\documentclass[%
 jmp,
 amsmath,amssymb,
 reprint,%
]{revtex4-1}

\usepackage{graphicx}% Include figure files
\usepackage{dcolumn}% Align table columns on decimal point
\usepackage{bm}% bold math
\usepackage[utf8]{inputenc}
\usepackage[T1]{fontenc}
\usepackage{mathptmx}
\usepackage{float} % Mike added in this package on 6/1/2020 for inserting the captions

\begin{document}

%\preprint{AIP/123-QED}

\title[Converting translation operators into plane polar and spherical coordinates $\cdots$]{Converting translation operators into plane polar and spherical coordinates and their use in determining quantum-mechanical wavefunctions in a representation-independent fashion}
% Force line breaks with \\

\author{Michael Rushka}
% \affiliation{Department of Physics, Georgetown University, 37th and O Sts. NW, Washington, DC 20057 USA.}%Lines break automatically or can be forced with \\
\author{Mark Esrick}%
%\affiliation{ 
%Department of Physics, Georgetown University, 37th and O Sts. NW, Washington, DC 20057 USA}%

\author{W. N. Mathews Jr.}
%\affiliation{%
%Department of Physics, Georgetown University, 37th and O Sts. NW, Washington, DC 20057 USA}%

\author{ J. K. Freericks}
\affiliation{%
Department of Physics, Georgetown University, 37th and O Sts. NW, Washington, DC 20057 USA
}%

\date{\today}% It is always \today, today,

\begin{abstract}
Quantum mechanics is often developed in the position representation, but this is not necessary, and one can perform calculations in a representation-independent fashion, even for wavefunctions. In this work, we illustrate how one can determine wavefunctions, aside from normalization, using only operators and how those operators act on state vectors. To do this in plane polar and spherical coordinates requires one to convert the translation operator into those coordinates. As examples of this approach, we illustrate the solution of the Coulomb problem in two and three dimensions without needing to express any operators in position space.
\end{abstract}

\maketitle

\section{Introduction}

The quantum-mechanical position-space translation operator 
\begin{equation}
    \widehat{T}(\vec{r})=e^{-\frac{i}{\hbar}\vec{r}\cdot\hat{\vec{p}}}=e^{-\frac{i}{\hbar} \left (r_x\hat{p}_x+r_y\hat{p}_y+r_z\hat{p}_z\right)},
    \label{eq: translation_op} 
\end{equation}
is normally expressed in terms of Cartesian components, 
\begin{equation}
\vec{r} = \vec{e}_x r_x + \vec{e}_y r_y + \vec{e}_z r_z, \; \hat{\vec{p}} = \vec{e}_x \hat{p}_x + \vec{e}_y \hat{p}_y + \vec{e}_z \hat{p}_z ,
\end{equation}
with $\vec{e}_x$, $\vec{e}_y$ and $\vec{e}_z$ the Cartesian unit vectors.
We use hats to denote operators on the state vector space.  Note, in particular that the quantities $r_\alpha$ in the translation operator are numbers, not operators.  The position and momentum operators satisfy the canonical commutation relation,
\begin{equation} 
 [\hat r_\alpha,\hat p_\beta]=i\hbar\delta_{\alpha,\beta} ,
\end{equation} 
where $\alpha$ and $\beta$ run over the Cartesian dimensions.  

The translation operator can be employed to determine position eigenstates by translating the position-space eigenvector at the origin, $|0_{\vec{r}} \rangle$, which satisfies
\begin{equation}
\hat{r}_{\alpha} |0_{\vec{r}} \rangle = 0.
\label{eq: origin}
\end{equation}
One of our assumptions is that such a position eigenvector at the origin {\it exists}, without worrying about the details of rigged Hilbert spaces; we will see that the wavefunction at any position can be determined relative to the wavefunction at the origin.  We then have the position-operator eigenstates given by
\begin{equation}
    |\vec{r}\rangle = \widehat{T}(\vec{r}) \; |0_{\vec{r}} \rangle .
    \label{eq: position_eigenvector}
\end{equation}
It is easy to verify that  
\begin{equation}
\hat r_{\alpha} |\vec{r} \rangle = r_{\alpha} |\vec{r} \rangle , \end{equation}
by using the braiding relation (derived from the Hadamard lemma) 
\begin{align}
e^{\hat{A}} f(\hat{B})e^{-\hat{A}} = &f \big( \hat{B} + [\hat{A},\hat{B}] + \frac{1}{2!}[\hat{A},[\hat{A},\hat{B}]]
+ \\ &+ \frac{1}{3!} [\hat{A},[\hat{A},[\hat{A},\hat{B}]]]+\cdots \big)  \notag, 
   \label{eq: hadamard}
\end{align}
which is valid provided $f(\hat{B})$ can be written as a sum on nonnegative integral powers of $\hat{B}$ and the argument of the function on the right hand side is a sum of terms consisting of increasingly nested commutators. For the verification of the position eigenstate, one simply notes that the commutator of position with momentum is a number and so it commutes with all operators, which truncates the series after the first commutator.  

The strategy of a representation-independent approach to quantum mechanics is to write the position-space wavefunction of a state $|\psi\rangle$ in terms of the position eigenstate at the origin and the translation operator, according to
\begin{equation} \label{eq: wavefunction_def}
    \psi(\vec{r})=\langle \vec{r}|\psi\rangle=\langle 0_{\vec{r}}|\widehat{T}^{\;\dagger}(\vec{r}) |\psi\rangle .
\end{equation}
 This expression provides a route to evaluate the wavefunction solely by manipulating operators. This  is because when the position operator acts on the position eigenstate at the origin it {\it annihilates} the state at the origin [see Eq.~(\ref{eq: origin})].  To carry out these calculations for energy eigenfunctions, one needs to factorize the Hamiltonian, in order to determine what happens when the momentum operator acts on the state $|\psi\rangle$, using the methodology of the Schr\"odinger factorization method~\cite{schroedinger,green,ohanian}.

We term this approach a representation-independent way to calculate the wavefunctions, because one does not need to express the operators used to determine the state vectors in the specific basis of the wavefunctions. Instead, we only employ the commutation relations of the operators to determine the wavefunctions (up to overall normalization). For example, when working with the momentum operator, we only use the canonical commutation relation and how the momentum operator acts on the energy eigenstate.  We do not need to use $\hat{\vec{p}}=-i\hbar\nabla$, the position-space representation of the momentum operator.

B\"ohm illustrated an alternate way to calculate wavefunctions in a representation-independent fashion~\cite{bohm}, and Merzbacher also used this approach~\cite{merzbacher}. Matrix elements of the position operator between energy eigenstates of the simple harmonic oscillator were employed to determine recurrence relations between energy eigenfunctions of different energy eigenvalues, but at the same position. The recurrence relations were then solved in terms of Hermite polynomials, eventually yielding the usual position-space wavefunctions. The general approach we develop here instead fixes the eigenfunction and relates the value of the wavefunction at the origin to the value of the wavefunction at $\vec{r}$. This latter approach can be generalized to many different problems other than the simple harmonic oscillator. It is not clear whether the method developed by B\"ohm can be extended beyond the simple harmonic oscillator.
Our approach can also be employed for momentum-space wavefunctions.

 We illustrate this process below.  But before doing so, we point out the reason why we need to transform the translation operator to spherical and plane polar coordinates. Spherically symmetric problems, such as the Coulomb problem, have wavefunctions that are explicit functions of 
\begin{equation} \label{eq: r}
r=\sqrt{r_x^2+r_y^2+r_z^2}.
\end{equation}
Such functions, for odd powers of $r$, cannot be expanded in a Maclauren series in the Cartesian position components, because expansions about the origin in position space do not exist. Hence, one cannot use the Cartesian-basis representation of the translation operator.  A similar argument applies for plane polar coordinates.

In Sec.~II, we provide the details of how to convert the translation operator from the expression in terms of Cartesian components of the momentum operator to an expression in terms of the components of the momentum operator in spherical and plane polar coordinates. The exact operator expression can be simplified when it acts on $|0_{\vec{r}}\rangle$, which is the final form we employ to calculate wavefunctions. In Sec. III, we employ this approach to calculate the position-space wavefunctions for two and three-dimensional Coulomb problems. This methodology is based on Schr\"odinger's factorization method. We conclude the paper in Sec. IV. In 
the Appendix, we summarize technical details, primarily related to computing commutators in a representation-independent fashion.

\section{Translation operator in spherical coordinates}

To work with a three-dimensional spherically symmetric system, we express the translation operator, $\widehat{T}(\vec{r})$, in terms 
of the spherical coordinates, $r$, $\theta$, $\phi$, the corresponding position operators, $\hat{r}$, $\cos\hat{\theta}$, $\sin\hat{\theta}$, $\cos\hat{\phi}$, $\sin\hat{\phi}$ and $\hat{p}_r$, $\hat{p}_{\theta}$, $\hat{p}_{\phi}$, the spherical components of the momentum operator, $\hat{\vec{p}}$. We use the spherical components instead of the canonical momenta, because this is the cleanest way to decompose the inner product $\vec{r}\cdot\hat{\vec{p}}$. Note that $\hat{p}_r$ is a canonical momentum operator because it is the quantum analog of the classical momentum conjugate to the radial degree of freedom, whereas the same is not true of $\hat{p}_{\theta}$ and $\hat{p}_{\phi}$.  We can still express the translation operator in terms of $\hat{p}_r$, $\hat{p}_{\theta}$, $\hat{p}_{\phi}$ even though $\hat{p}_r$ is  Hermitian, but not self adjoint, and thus is not an observable. In addition, the translation operator remains unitary regardless of the coordinate system in terms of which it is expressed.

There is a subtle point associated with the symbol $|0_{\vec{r}} \rangle$:
while $\vec{r} = 0$ implies $r_x = r_y = r_z = 0$, it only implies $r = 0$ in spherical coordinates. That is, the values of $\theta$ and $\phi$ are indeterminate until specified by some limiting procedure for how the origin is approached.  We will use what we call a ``north-pole'' state oriented along the positive $z$-axis for definiteness.

It is well  known that defining operators corresponding to the angles $\hat{\theta}$ and $\hat{\phi}$ is problematic~\cite{phase_review}. Instead, we define the values corresponding to the position eigenstates expressed in spherical coordinates through the cosine or sine of those operators, which are always well defined in terms of $\hat{r}_\alpha$.  In particular,
\begin{equation}
\cos \hat{\theta} = \frac{\hat{r_z}}{\hat{r}} , \; \sin \hat{\theta} = \frac{\hat{\rho}}{\hat{r}} , \; \cos \hat{\phi} = \frac{\hat{r}_x}{\hat{\rho}} , \; \sin \hat{\phi} = \frac{\hat{r}_y}{\hat{\rho}} ,
\label{eq: trig_def}
\end{equation}
where
\begin{equation}
\hat{r}=\sqrt{\hat{r}_x^{\;2} + \hat{r}_y^{\;2}+ \hat{r}_z^{\;2}}~~\text{and}~~\hat{\rho} = \sqrt{\hat{r}_x^{\;2} + \hat{r}_y^{\;2}}.
\label{eq: radial_def}
\end{equation}
These are all well defined, but care must be taken when these operators act on position eigenstates that approach the origin.

\begin{figure}[bth]
\centerline{\includegraphics[width=1.9in]{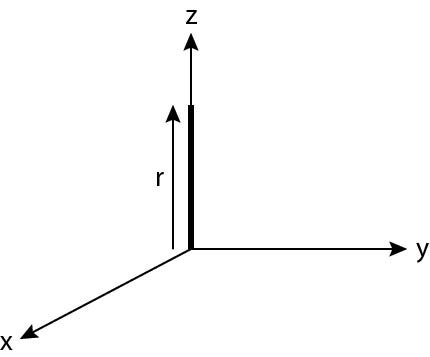}}
\centerline{\includegraphics[width=1.9in]{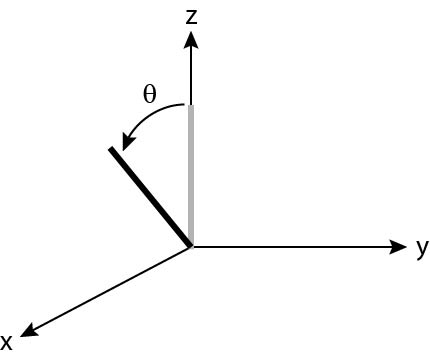}}
\centerline{\includegraphics[width=1.9in]{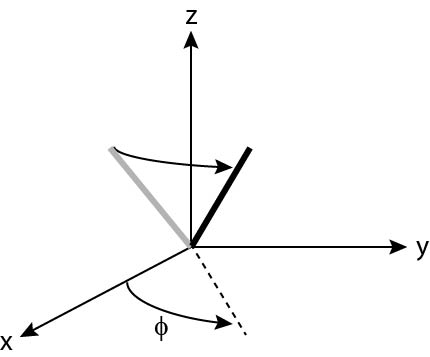}}
\caption{Three-step process to move from the origin to $(r_x,r_y,r_z)$: (top) first translate a distance $r$ along the $z$-axis; (middle) rotate about the $y$-axis by an angle $\theta$; and (bottom) rotate by an angle $\phi$ about the $z$-axis. One can also reach the same point by rotating directly after the translation in the $z$-direction to the final point, or one can translate in the $\theta$-direction in the $x-z$ plane and then rotate by $\phi$, or one can translate in the $\theta$, $\phi$ direction a distance $r$ directly (not shown). All these alternatives lead to the same final point in space. The operators corresponding to each of these different ways to translate and rotate from the origin to the final point are shown in the main text.}
\end{figure}

Arguably, the easiest way to go from $|0_{\vec{r}}\rangle$ to $|\vec{r}\rangle$ using spherical coordinates is to first  translate a distance $r$ in the $z$-direction, rotate by $\theta$ about the $y$-axis, and then rotate by $\phi$ about the $z$-axis, as illustrated in Fig. 1.  Indeed, this is the common way to define the spherical coordinates $r$, $\theta$ and $\phi$.
The operator needed to do this is 
\begin{equation} \label{eq: r, theta, phi route}
\widehat{T}(\vec{r}) = e^{- \frac{i}{\hbar} \phi\hat{L}_z} e^{- \frac{i}{\hbar} \theta\hat{L}_y} e^{-\frac{i}{\hbar} r\hat{p}_z} e^{  \frac{i}{\hbar} \theta\hat{L}_y} e^{  \frac{i}{\hbar}\phi \hat{L}_z} ,
\end{equation}
and we discuss below the subtleties associated with how it operates on $|0_{\vec{r}}\rangle$.  Here $\hat{L}_y$ and $\hat{L}_z$ are the $y$- and $z$- components, respectively, of $\hat{\vec{L}}=\hat{\vec{r}}\times\hat{\vec{p}}$, the orbital angular momentum operator. 

To establish this result, we use the braiding relation twice. First we note that
\begin{equation}
e^{- \frac{i}{\hbar} \theta\hat{L}_y} e^{- \frac{i}{\hbar} r\hat{p}_z} e^{  \frac{i}{\hbar}\theta \hat{L}_y} = e^{- \frac{i}{\hbar} r(\sin \theta \hat{p}_x +  \cos \theta\hat{p}_z)}
\end{equation}
and then
\begin{eqnarray} \label{eq: intermediate T} 
\widehat{T}(\vec{r}) &=&  e^{- \frac{i}{\hbar}\phi\hat{L}_z}e^{- \frac{i}{\hbar} r( \sin \theta\hat{p}_x + \cos \theta\hat{p}_z )} e^{- \frac{i}{\hbar} \phi\hat{L}_z}\\
&=& e^{- \frac{i}{\hbar}r (\sin \theta \cos \phi \: \hat{p}_x + \sin \theta \sin \phi \:\hat{p}_y + \cos \theta \: \hat{p}_z)}.\nonumber
\end{eqnarray}
Since 
\begin{equation}
r_x=r\sin\theta\cos\phi,~ r_y=r\sin\theta\sin\phi~~\text{and}~~ r_z=r\cos\theta ,
\end{equation}
Eq. \eqref{eq: intermediate T} obviously yields the translation operator as expressed in Cartesian coordinates in Eq. \eqref{eq: translation_op}. 

We use Eqs.~\eqref{eq: radial_mom} and \eqref{eq: theta_mom} to express $\hat{p}_z$ in terms of the spherical components of momentum according to
\begin{equation} \label{eq: p sub z}
\hat{p}_z =  \Big(\hat{p}_r - i \frac{\hbar}{2 \hat{r}}\Big)\cos \hat{\theta} -  \hat{p}_{\theta} \sin \hat{\theta}.
\end{equation}
Note that the ordering is important in the second term because $\sin\hat{\theta}$ does not commute with $\hat{p}_\theta$, but there is no ordering ambiguity with the first term.
Equations \eqref{eq: r, theta, phi route} and \eqref{eq: p sub z} allow us to eliminate the Cartesian components of momentum from the translation operator and obtain
\begin{eqnarray}
\widehat{T}(\vec{r}) &=& e^{ - \frac{i}{\hbar} \phi \hat{L}_z} e^{ - \frac{i}{\hbar} \theta \hat{L}_y} \times \\ &\times& e^{- \frac{i}{\hbar} r  \big[ \big(\hat{p}_r - i \frac{\hbar}{2 \hat{r}} \big) \cos \hat{\theta} -  \hat{p}_{\theta}\;\sin \hat{\theta} \big]} e^{ \frac{i}{\hbar} \theta \hat{L}_y} e^{  \frac{i}{\hbar} \phi \hat{L}_z} \notag .
\end{eqnarray}
We use Eq.~\eqref{eq: theta_mom} and 
\begin{equation} \label{eq:e sub phi}
\hat{\vec{e}}_{\phi} = -\vec{e}_x \sin \hat{\phi} + \vec{e}_y \cos \hat{\phi} ,
\end{equation}
to re-express $\hat{p}_{\theta}$ according to
\begin{equation}
\hat{p}_{\theta} = \frac{1}{\hat{r}} \Big(\hat{\vec{L}}\;\cdot\;\hat{\vec{e}}_{\phi} + i \frac{\hbar}{2} \cot \hat{\theta} \Big),
\end{equation}
and thus transform the translation operator into its final form
\begin{eqnarray} \label{eq: translation_op 3}
\widehat{T}(\vec{r}) &=& e^{- \frac{i}{\hbar} \phi \hat{L}_z} e^{- \frac{i}{\hbar} \theta \hat{L}_y} \times \\ &\times&e^{- \frac{i}{\hbar} r \big[\big(\hat{p}_r - i \frac{\hbar}{\hat{r}} \big)\cos \hat{\theta}  - \frac{\hat{\vec{L}}\;\cdot\;\hat{\vec{e}}_{\phi}}{\hat{r}} \sin \hat{\theta} \big]} e^{  \frac{i}{\hbar} \theta \hat{L}_y} e^{  \frac{i}{\hbar} \phi \hat{L}_z} \notag.
\end{eqnarray}
We emphasize that this is an operator equality. It is the expression of the translation operator in terms of spherical coordinates.

This operator relation can be rewritten in three other forms, which illustrate the different ways that one can derive the translation operator in spherical coordinates. First, note that $\hat L_z$ commutes with $\hat r$, $\cos\hat\theta$, $\sin\hat\theta$, $\hat{\vec{L}}\cdot\hat{\vec{e}}_\phi$, and $\hat{p}_r$ (or, more simply, it commutes with $\hat{p}_z$). This means we can introduce the factor $\exp\left (\frac{i}{\hbar}\phi\hat{L}_z\right)\exp\left ( - \frac{i}{\hbar}\phi\hat{L}_z\right)$ just to the left of the $\exp\left (\frac{i}{\hbar}\theta\hat{L}_y\right )$ term in Eq.~(\ref{eq: translation_op 3}) and then move the factor $\exp\left (\frac{i}{\hbar}\phi\hat{L}_z\right)$ to the left through the middle exponential factor yielding
\begin{eqnarray} \label{eq: translation_op 4}
\widehat{T}(\vec{r}) &=& e^{- \frac{i}{\hbar} \phi \hat{L}_z} e^{- \frac{i}{\hbar} \theta \hat{L}_y} e^{ \frac{i}{\hbar} \phi \hat{L}_z}\times \\ &\times&e^{- \frac{i}{\hbar} r \big[\big(\hat{p}_r - i \frac{\hbar}{\hat{r}} \big)\cos \hat{\theta}  - \frac{\hat{\vec{L}}\;\cdot\;\hat{\vec{e}}_{\phi}}{\hat{r}} \sin \hat{\theta} \big]} e^{- \frac{i}{\hbar} \phi \hat{L}_z}e^{  \frac{i}{\hbar} \theta \hat{L}_y} e^{  \frac{i}{\hbar} \phi \hat{L}_z} \notag.
\end{eqnarray}
This form of the translation operator expresses it as a similarity transformation of the  middle exponential factor  with respect to the operator
\begin{align}
 e^{ - \frac{i}{\hbar} \phi \hat{L}_z}  e^{ - \frac{i}{\hbar} \theta \hat{L}_y} e^{ \frac{i}{\hbar} \phi \hat{L}_z} &= e^{ - \frac{i}{\hbar} \theta ( -  \hat{L}_x \sin \phi + \hat{L}_y \cos \phi )} \\ &= e^{ - \frac{i}{\hbar} \theta \vec{e}_{\phi}\;\cdot\;\hat{\vec{L}}},
\end{align}
which follows from the braiding relation. Note that
\begin{equation}
\vec{e}_{\phi}\;\cdot\;\hat{\vec{L}} =  -  \hat{L}_x \sin \phi +  \hat{L}_y \cos \phi 
\end{equation}
is a linear combination of the angular momentum operators with numbers, not operators, as coefficients because $\vec{e}_\phi$ is {\it not} an operator here. The translation operator then becomes
\begin{equation}
    \widehat{T}(\vec{r}) = e^{ - \frac{i}{\hbar} \theta \vec{e}_{\phi}\;\cdot\;\hat{\vec{L}}}
    e^{- \frac{i}{\hbar} r \big[\big(\hat{p}_r - i \frac{\hbar}{\hat{r}} \big)\cos \hat{\theta}  - \frac{\hat{\vec{L}}\;\cdot\;\hat{\vec{e}}_{\phi}}{\hat{r}} \sin \hat{\theta} \big]}
    e^{ \frac{i}{\hbar} \theta \vec{e}_{\phi}\;\cdot\;\hat{\vec{L}}}.
\end{equation}
In this form of the translation operator, we first translate along the $z$-axis a distance $r$ and then rotate by an angle $\theta$ about an axis along $\vec{e}_{\phi}$, i.e., an axis rotated an angle $\phi$ counterclockwise from the $y$-axis.

The final two ways we express the operator come from a simple brute-force substitution. We solve Eqs. \eqref{eq: radial_mom}, \eqref{eq: theta_mom}, and \eqref{eq: phi_mom} for $\hat{p}_x$, $\hat{p}_y$, and $\hat{p}_z$
 in terms of $\hat{p}_r$, $\hat{p}_{\theta}$, and $\hat{p}_{\phi}$ and substitute into Eq. \eqref{eq: translation_op}.  Then we use the inverse of the braiding relation to remove an $\exp\left (-\frac{i}{\hbar} \theta \hat{L}_z\right )$ to the left and its hermitian conjugate to the right. We also obtain this form by using the braiding relation to move the $\exp\left (-\frac{i}{\hbar}\theta\hat{L}_y\right )$ factor (and its conjugate) into the exponent in Eq.~(\ref{eq: translation_op 3}). This yields
\begin{align}
\label{eq: translation_op 5}
\widehat{T}(\vec{r}) &= e^{ - i \frac{\phi}{\hbar} \hat{L}_z} \exp \Bigg( -  \frac{i}{\hbar}r \Bigg\{ \Big(\hat{p}_r - i \frac{\hbar}{\hat{r}} \Big) \cos (\hat{\theta} - \theta) - \\ &- \frac{1}{\hat{r}} \hat{\vec{L}} \cdot \hat{\vec{e}}_{\phi} \sin(\hat{\theta}-\theta) + \Big[ \Big(\hat{p}_r -i \frac{\hbar}{\hat{r}} \Big) \sin \theta \sin \hat{\theta} + \notag \\ &+ \frac{1}{\hat{r}} \hat{\vec{L}} \cdot \hat{\vec{e}}_{\phi} \sin \theta \cos \hat{\theta} \Big] (\cos \hat{\phi} - 1) -  \frac{\hat{L}_z}{\hat{r}}\frac{\sin \theta \sin \hat{\phi}}{\sin \hat{\theta}} \Bigg\} \Bigg) e^{ i \frac{\phi}{\hbar} \hat{L \notag
 }_z} .
\end{align}
One can interpret this as a translation along the $\theta$ direction in the $x-z$ plane, followed by a rotation by $\phi$ about the $z$-axis. If we use the braiding operation to move the remaining exponential factor into the exponent, we would have a single translation of length $r$ along the $\theta$, $\phi$ direction. We do not write that final other form explicitly here; it is given by the result in Eq.~(\ref{eq: translation_op 5}) without the exponential factors on the left and the right and with the substitution $\hat\phi\to \hat\phi-\phi$.

As one can immediately see, these forms for the translation operator are equivalent and are easily related by employing the braiding relation in different ways. They also have different physical interpretations for how the translation is performed. Of course the operators in all four cases are identical, and are just expressed in terms of different exponential factors.

The remaining task is to determine the proper limiting procedure to apply this operator on the position eigenstate at the origin. We derive it explicitly for the form of the translation operator given in Eq.~(\ref{eq: translation_op 3}), but the generalization for any of the other forms is straightforward to work out. Our final result is independent of which form is used.

The strategy is to perform a translation along $\vec{r}$, so that the indeterminate operators corresponding to $\cos \hat{\theta}$ and $\sin \hat{\theta}$ [when acting on $|0_{\vec r}\rangle$] can be properly defined.  [It turns out that the indeterminacy with respect to $\phi$ need not be resolved in order to determine the final formula for the translation operator.]  To this end, we introduce $\exp\left ({ \frac{i}{\hbar} \vec{\delta}\;\cdot \; \hat{\vec{p}}}\right ) \exp\left ({ -  \frac{i}{\hbar} \vec{\delta}\;\cdot \; \hat{\vec{p}}}\right ) = 1$ to the left of $\widehat{T}(\vec{r})$.  The vector $\vec{\delta}$ is a number, not an operator, and we choose it to be in the same direction as the vector $\vec{r}$ in the translation operator. We then move the rightmost exponential factor through $\widehat{T}(\vec{r})$ to the right, which we can do since it commutes with $\widehat{T}(\vec{r})$, as is easy to see when $\widehat{T}(\vec{r})$ is expressed in its Cartesian form in Eq.~(\ref{eq: translation_op}).  We then operate on $|0_{\vec{r}}\rangle$, which yields
\begin{equation} \label{eq:Tran1}
|\vec{r}\rangle = e^{ \frac{i}{\hbar} \vec{\delta}\;\cdot\;\hat{\vec{p}}} \widehat{T}(\vec{r}) e^{ -  \frac{i}{\hbar} \vec{\delta}\;\cdot\;\hat{\vec{p}}} |0_{\vec{r}} \rangle .
\end{equation}
This equation can be re-expressed as
\begin{equation} \label{eq:Tran2}
|\vec{r}\rangle = e^{\frac{i}{\hbar} \vec{\delta}\;\cdot\;\hat{\vec{p}}} \widehat{T}(\vec{r}) |\delta_x, \delta_y, \delta_z \rangle .
\end{equation}
Since the two rotation operators on the right hand side of Eq.~(\ref{eq: translation_op 3}) first rotate by an angle $-\phi$ about the $z$-axis and then by an angle $-\theta$ about the $y$-axis, we find that the final position eigenstate, after the two rotations,  points along the $+ z$-axis. In other words, the choice that $\vec{\delta}$ lies in the same direction as $\vec{r}$, implies that
\begin{equation} \label{eq:phi cond}
\tan \phi = \frac{\delta_y}{\delta_x},
\end{equation}
so that
\begin{equation}
e^{  \frac{i}{\hbar} \phi \hat{L}_z} |\delta_x, \delta_y, \delta_z \rangle = |\sqrt{\delta_x^2 + \delta_y^2}, 0, \delta_z \rangle ,
\end{equation}
and implies further that
\begin{equation}
\tan \theta = \frac{\sqrt{\delta_x^2 + \delta_y^2}}{\delta_z},
\end{equation}
so that
\begin{equation}
e^{ \frac{i}{\hbar} \theta \hat{L}_y} e^{  \frac{i}{\hbar} \phi \hat{L}_z} |\delta_x, \delta_y, \delta_z \rangle = |0, 0, \delta \rangle .
\end{equation}
Here
\begin{equation} \label{eq:delta}
\delta=\sqrt{\delta_x^2 + \delta_y^2 + \delta_z^2}.
\end{equation}
Note that the ket $|0,0,\delta\rangle$ satisfies 
\begin{equation} \label{eq: cosine_action}
\cos\hat\theta|0, 0, \delta\rangle = \frac{\hat{r}_z}{\hat{r}}|0, 0, \delta\rangle\\
= |0, 0, \delta\rangle
\end{equation}
and 
\begin{equation}
\label{eq: sine_action}
\sin\hat\theta|0, 0, \delta\rangle=\frac{\hat{\rho}}{\hat r}|0, 0, \delta\rangle=0.
\end{equation}
Because this is a state oriented along the north pole, the action of $\cos\hat\phi$ or $\sin\hat\phi$ on this ket is indeterminate, in the sense that one cannot determine the action of $\cos\hat\phi$ or $\sin\hat\phi$ on this state.  
 
Our goal is to simplify the form of the translation operator when it acts on this ``north pole'' state. We expand the exponential function $\exp\left \{-\frac{i}{\hbar}\left [\left (\hat{p}_r-i\frac{\hbar}{\hat r}\right)\cos\hat\theta-\frac{\hat{\vec{L}}\cdot\hat{\vec{e}}_\phi}{\hat r}\sin\hat\theta\right ]\right \}$ in a power series and use the action of $\cos\hat\theta$ and $\sin\hat\theta$ on the ``north-pole'' state [in Eqs.~(\ref{eq: cosine_action}) and (\ref{eq: sine_action})]. Term by term in the power-series expansion, we see that the exponential function simplifies and can be resummed to the form $\exp\left [-\frac{i}{\hbar}\left (\hat{p}_r-i\frac{\hbar}{\hat r}\right)\right ]$ acting on the ``north-pole'' state. This produces 
\begin{equation}
|\vec{r}\rangle = e^{ \frac{i}{\hbar}\vec{\delta}\cdot\hat{\vec{p}}} e^{- \frac{i}{\hbar} \phi \hat{L}_z} e^{- \frac{i}{\hbar} \theta \hat{L}_y} e^{- \frac{i}{\hbar} r \big(\hat{p}_r - i \frac{\hbar}{\hat{r}} \big)} |0,0,\delta \rangle ,
\end{equation}

At this point, because $\hat{\vec{L}}$ commutes with $\hat{r}$ and $\hat{p}_r$, we can separate the radial and angular degrees of freedom according to
\begin{equation}  \label{eq: separation}
|\delta_x,\delta_y,\delta_z\rangle=e^{- \frac{i}{\hbar} \phi \hat{L}_z} e^{- \frac{i}{\hbar} \theta \hat{L}_y} |0,0,\delta \rangle = |r{=}\delta\rangle \otimes |\theta,\phi\rangle.
\end{equation} 
This result for the label of the state with $r{=}\delta$ arises because we define the radial coordinate eigenstates, $|r\rangle$, to satisfy 
\begin{equation}
\hat r|r\rangle = r |r\rangle ,
\end{equation}
with $r$ given by Eq. \eqref{eq: r}.  One can immediately verify that $\hat r|\delta_x,\delta_y,\delta_z\rangle=\delta|\delta_x,\delta_y,\delta_z\rangle$, which establishes the use of the label $r{=}\delta$ in Eq.~(\ref{eq: separation}).

We similarly define the angular state, $|\theta, \phi \rangle$, to be the state that satisfies
\begin{align}
\label{eq: theta_op_def}
\cos\hat\theta|\theta,\phi\rangle &= \cos\theta|\theta,\phi\rangle, \sin\hat\theta|\theta,\phi\rangle = \sin\theta|\theta,\phi\rangle,  \\  \cos\hat\phi|\theta,\phi\rangle &= \cos\phi|\theta,\phi\rangle, \sin\hat\phi|\theta,\phi\rangle = \sin\phi|\theta,\phi\rangle, 
\label{eq: phi_op_def}
\end{align}
with one exception. The eigenvalue-eigenvector relations in Eq.~(\ref{eq: phi_op_def}) cannot be satisfied when the eigenvalues in Eq.~(\ref{eq: theta_op_def}) correspond to the cases where $\theta=0$ or $\theta=\pi$.

Note that the operators $\hat{r}_x$, $\hat{r}_y$ and $\hat{r}_z$ cannot operate solely on the state $|r\rangle$, i.e., the domain of those operators lies outside of the space of the eigenstates of the radial-position operator $\hat{r}$.  Similarly, they cannot act on the state $|\theta,\phi\rangle$. Of course, they can act on the tensor-product state $|r\rangle\otimes|\theta,\phi\rangle$. 

However, we can allow the operator $\hat r$ to act only on $|r\rangle$ (and as the identity operator on $|\theta,\phi\rangle$) and $\cos\hat\theta$ (and the other similar trigonometric operators) act only on $|\theta,\phi\rangle$ (and as the identity on $|r\rangle$) as we see next.
Consider $\cos\hat\theta$ acting on the state $|r\sin\theta\cos\phi,r\sin\theta\sin\phi,r\cos\theta\rangle=|r\rangle\otimes|\theta,\phi\rangle$, which is given by
\begin{align}
    \cos\hat\theta |r\rangle \otimes |\theta,\phi\rangle&=\frac{\hat{r}_z}{\hat{r}}|r\rangle \otimes |\theta,\phi\rangle\\
    &=e^{- \frac{i}{\hbar} \phi \hat{L}_z}\hat{r}_z e^{- \frac{i}{\hbar} \theta \hat{L}_y}\frac{1}{\hat r} |0,0,r \rangle\notag\\
    &=e^{- \frac{i}{\hbar} \phi \hat{L}_z}e^{- \frac{i}{\hbar} \theta \hat{L}_y}\underbrace{e^{ \frac{i}{\hbar} \theta \hat{L}_y}\hat{r}_ze^{- \frac{i}{\hbar} \theta \hat{L}_y}}_{\text{braiding}}\frac{1}{\hat r} |0,0,r \rangle\notag\\
    &=e^{- \frac{i}{\hbar} \phi \hat{L}_z}e^{- \frac{i}{\hbar} \theta \hat{L}_y}\frac{ - \sin\theta\hat{r}_x+\cos\theta\hat{r}_z}{\hat r} |0,0,r \rangle\notag\\ 
     &=\cos\theta |r\rangle \otimes |\theta,\phi\rangle.\notag
\end{align}
We can accomplish this because the eigenvalue-eigenvector relationship given above is unchanged when we change the value of $r$ in the radial ket. Hence, the operator $\cos\hat\theta$ acts as the identity on the $|r\rangle$ ket and can be taken to act solely on the $|\theta,\phi\rangle$ ket, yielding $\cos\hat\theta|\theta,\phi\rangle=\cos\theta|\theta,\phi\rangle$. We can proceed similarly to verify that this separation holds true for the three other trigonometric operators. The argument for the radial operator acting only on $|r\rangle$ can also be easily verified.

We can then take the limit $\delta \to 0^+$, so that $\exp\left (\frac{i}{\hbar}\vec{\delta}\cdot\hat{\vec{p}}\right) \to 1$.
Gathering the final results together, we have established that
\begin{equation} 
\label{eq:final ket}
|\vec{r} \rangle = e^{- \frac{i}{\hbar} r \big(\hat{p}_r - i \frac{\hbar}{\hat{r}} \big)} |r{=}0\rangle\otimes |\theta,\phi\rangle
\end{equation}
and
\begin{equation} 
\label{eq:final bra}
\langle \vec{r} | =\langle \theta, \phi |\otimes\langle r{=}0|e^{  \frac{i}{\hbar}r \big({p}_r + i \frac{\hbar}{\hat{r}} \big)}  .
\end{equation}
These final expressions are a simplification of the translation operator in spherical coordinates when it acts on the state at the origin.

In order to calculate an energy wavefunction in position space, we must also decompose the energy eigenstate vector of the Hamiltonian according to a radial and angular-momentum tensor-product state via
\begin{equation}
|\Psi \rangle = |\psi_r \rangle \otimes |l,m \rangle ,  
\end{equation}
where the eigenvalues of $\hat{\vec{L}}^2$ and ${\hat{L}}_z$ (when acting on the state $|l,m\rangle$) are $l (l + 1) \hbar^2$ and $m \hbar$, respectively.  It follows that we can write the wave function as the product of the radial wavefunction and the angular momentum eigenfunction according to
\begin{equation}
\langle \vec{r}|\Psi \rangle = \langle r | \psi_r \rangle \langle \theta, \phi |l, m \rangle
\end{equation}
and
\begin{equation}
\Psi (\vec{r}) = \psi_r(r) Y_{l ,m}(\theta,\phi) ,
\end{equation}
where
\begin{equation}
Y_{l , m} (\theta, \phi) = \langle \theta,\phi|l,m\rangle
\end{equation}
denotes the standard spherical harmonic.  A derivation of the spherical harmonic in a representation-independent way using the rotation operators of the ``north-pole'' state can be found in Ref.~\cite{weitzman_freericks}.  The radial wavefunction is thus given by
\begin{equation}
\psi_r(r) = \langle r |\psi_r \rangle = \langle r{=}0 |e^{ \frac{i}{\hbar}r \big({p}_r + i \frac{\hbar}{\hat{r}} \big)}|\psi_r \rangle .
\end{equation}
This is the relation that we will use to determine the radial wavefunctions for the three-dimensional Hydrogen atom. 

We now turn to the expression of the two-dimensional translation operator in plane polar coordinates.  In terms of Cartesian coordinates, we have
\begin{equation} \label{eq: 2D translation op 1} 
    \widehat{T}(\vec{\rho})=e^{-\frac{i}{\hbar}\vec{\rho}\cdot\hat{\vec{p}}}=e^{-\frac{i}{\hbar} \left (r_x\hat{p}_x+r_y\hat{p}_y\right)},
\end{equation}
with, 
\begin{equation}
\vec{\rho} = \vec{e}_x r_x + \vec{e}_y r_y , \; \hat{\vec{p}} = \vec{e}_x \hat{p}_x + \vec{e}_y \hat{p}_y.
\end{equation}
We use $|0_{\vec{\rho}}\rangle$ as the position-space eigenvector at the origin for this two-dimensional case.  We thus have
\begin{equation}
    |\vec{\rho}\rangle = \widehat{T}(\vec{\rho}) \; |0_{\vec{\rho}} \rangle .
    \label{eq: position_eigenvector}
\end{equation}

Probably the simplest way to go from $|0_{\vec{\rho}}\rangle$ to $|\vec{\rho}\rangle$ is to translate by $\rho$ in the $+x$-direction and then rotate by $\phi$ about the $+z$-direction.  Indeed, this is the usual way of defining plane polar coordinates, $\rho$ and $\phi$.  The corresponding operator is
\begin{equation} \label{eq: rho, phi route}
\widehat{T}(\vec{\rho}) = e^{- \frac{i}{\hbar} \phi\hat{L}_z}  e^{-\frac{i}{\hbar} \rho\hat{p}_x}  e^{  \frac{i}{\hbar}\phi \hat{L}_z} ,\
\end{equation}
To show that this is correct, we use the braiding relation to obtain
\begin{equation} \label{eq: 2D translation op 2}
\widehat{T}(\vec{\rho}) = e^{ - \frac{i}{\hbar} \rho (\hat{p}_x \cos \phi + \hat{p}_y \sin \phi )} .   
\end{equation}
Since
\begin{equation} \label{eq:Cartesian to plane polar}
r_x = \rho \cos \phi ~~\mathrm{and}~~ r_y = \rho \sin \phi    ,
\end{equation}
we see immediately that Eq.~\eqref{eq: 2D translation op 2} reduces to Eq.~\eqref{eq: 2D translation op 1}.

From Eqs.~\eqref{eq: radial_momentum_2d} and \eqref{eq: phi_momentum_2d}, we obtain
\begin{equation} \label{eq:p_x in 2D}
\hat{p}_x = \Big(\hat{p}_{\rho} - i \frac{\hbar}{2\;\hat{\rho}} \Big) \cos \hat{\phi} - \hat{p}_{\phi} \sin \hat{\phi} 
\end{equation}
and
\begin{equation} \label{eq:p_y in 2D}
\hat{p}_y = \Big(\hat{p}_{\rho} - i \frac{\hbar}{2\;\hat{\rho}} \Big) \sin \hat{\phi} + \hat{p}_{\phi} \cos \hat{\phi} .
\end{equation}
We use Eqs.~\eqref{eq: rho, phi route} and \eqref{eq:p_x in 2D} to write
\begin{equation} \label{eq: 2D translation op 3}
\widehat{T}(\vec{\rho}) = e^{- \frac{i}{\hbar} \phi\hat{L}_z}  e^{-\frac{i}{\hbar} \rho\Big[\Big(\hat{p}_{\rho} - i \frac{\hbar}{2\;\hat{\rho}} \Big) \cos \hat{\phi} - \hat{p}_{\phi} \sin \hat{\phi}\Big] }  e^{  \frac{i}{\hbar}\phi \hat{L}_z} .
\end{equation}
We note that this is an operator identity.

A second way to express $\widehat{T}(\vec{\rho})$ in plane polar coordinates is simply to use Eqs.~\eqref{eq:Cartesian to plane polar}, \eqref{eq:p_x in 2D}, and \eqref{eq:p_y in 2D} to substitute for the Cartesian quantities in Eq.~\eqref{eq: 2D translation op 1}.  In this way, we obtain
\begin{align}
r_x \hat{p}_x + r_y \hat{p}_y &= \rho \Big[ \Big( \hat{p}_r - i \frac{\hbar}{2\;\hat{\rho}}\Big) (\cos \phi \cos \hat{\phi} + \sin \phi \sin \hat{\phi}) - \\ &- \hat{p}_{\phi} (\cos \phi \sin \hat{\phi} - \sin \phi \cos \hat{\phi} )\Big]  \notag ,
\end{align}
\begin{equation}
r_x \hat{p}_x + r_y \hat{p}_y = \rho \Big[ \Big( \hat{p}_r - i \frac{\hbar}{2\;\hat{\rho}}\Big) \cos (\hat{\phi} - \phi) - \hat{p}_{\phi} \sin (\hat{\phi} -\phi)\Big]  ,
\end{equation}
and
\begin{equation}
r_x \hat{p}_x + r_y \hat{p}_y = e^{ - \frac{i}{\hbar} \phi \hat{L}_z} \rho \Big[ \Big( \hat{p}_r - i \frac{\hbar}{2\;\hat{\rho}}\Big) \cos \hat{\phi} - \hat{p}_{\phi} \sin \hat{\phi}\Big]   e^{ \frac{i}{\hbar} \phi \hat{L}_z}.
\end{equation}
Upon substitution of this into Eq. \eqref{eq: 2D translation op 1}, we immediately obtain Eq. \eqref{eq: 2D translation op 3}.

Next, we determine how the general operator form of the translation operator in plane polar coordinates simplifies, when it acts on the origin state in two dimensions,  $|0_{\vec\rho}\rangle$, similar to what we did in the three-dimensional case. 

We introduce the same pair of $\vec{\delta}$-dependent exponentials to the left of $\widehat{T}(\vec{\rho})$, where $\vec{\delta}$ is a two-dimensional vector along $\vec{\rho}$, move the rightmost exponential factor to the right through $\widehat{T}(\vec{\rho})$, and operate on $|0_{\vec{\rho}}\rangle$.  We thus obtain
\begin{equation} \label{eq:2D tran 1}
|\vec{\rho}\rangle = e^{\frac{i}{\hbar} \vec{\delta}\;\cdot\;\hat{\vec{p}}} \widehat{T}(\vec{\rho}) |\delta_x, \delta_y\rangle .
\end{equation}
The choice that $\vec{\delta}$ lies along the same direction as $\vec{\rho}$, implies that Eq.~\eqref{eq:phi cond} holds and
\begin{equation}
e^{\frac{i}{\hbar} \phi \hat{L}_z} |\delta_x, \delta_y \rangle = |\delta, 0 \rangle, 
\end{equation}
where $\delta$ is given by Eq.~\eqref{eq:delta} with $\delta_z = 0$.  Note that
\begin{equation}
\cos \hat{\phi}|\delta,0\rangle = \frac{\hat{r}_x}{\hat{\rho}}|\delta,0\rangle =  |\delta,0\rangle  
\end{equation}
and
\begin{equation}
\sin \hat{\phi}|\delta,0\rangle = \frac{\hat{r}_y}{\hat\rho}|\delta,0\rangle = 0 .
\end{equation}
We expand the exponential containing $\hat{p}_{\rho}$ in Eq. \eqref{eq: 2D translation op 3} and use the action of $\cos \hat{\phi}$ and $\sin \hat{\phi}$ on $|\delta,0\rangle$ to simplify each term and then re-sum to obtain $\exp\Big[ - \frac{i}{\hbar} \rho \Big(\hat{p}_{\rho} - i \frac{\hbar}{2\;\hat{\rho}} \Big) \Big]$ acting on the ket.  This yields
\begin{equation} \label{eq:2D tran 2}
|\vec{\rho}\rangle = e^{\frac{i}{\hbar} \vec{\delta}\;\cdot\;\hat{\vec{p}}} e^{ - \frac{i}{\hbar} \phi \hat{L}_z} e^{ - \frac{i}{\hbar} \rho \Big(\hat{p}_{\rho} - i \frac{\hbar}{2\;\hat{\rho}} \Big)}|\delta,0\rangle .
\end{equation}

Since $\hat{L}_z$ commutes with $\hat{p}_{\rho}$ and $\hat{\rho}$, we can separate the radial and angular degrees of freedom according to
\begin{equation} \label{eq:2D separation}
|\delta_x,\delta_y\rangle = e^{ - \frac{i}{\hbar} \phi \hat{L}_z} |\delta,0\rangle = |\rho{=} \delta\rangle \otimes |\phi\rangle .
\end{equation}
We define the radial coordinate eigenstate to satisfy
\begin{equation}
\hat{\rho}|\rho\rangle = \rho |\rho\rangle.
\end{equation}
A direct calculation gives $\hat{\rho} |\delta_x,\delta_y\rangle = \delta |\delta_x,\delta_y\rangle$ justifying the label $\rho = \delta$ in Eq.~\eqref{eq:2D separation}.
We similarly define the angular state, $|\phi \rangle$, to be the state that satisfies
\begin{equation}
\label{eq: phi op def}
\cos\hat\phi|\phi\rangle = \cos\phi|\phi\rangle, ~~\sin\hat\phi|\phi\rangle = \sin\phi|\phi\rangle.
\end{equation}

Note that the operators $\hat{r}_x$ and $\hat{r}_y$ cannot operate solely on the states $|\rho\rangle$ or $|\phi\rangle$; that is, the domain of those operators lies outside of the space of the eigenstates of the radial-position operator $\hat{\rho}$ and of the angle ket $|\phi\rangle$.  Of course, they can act on the tensor-product state $|\rho\rangle\otimes|\phi\rangle$ by decomposing the Cartesian position operators into their polar coordinate counterparts. 

But, the operator $\hat{\rho}$ does act only on $|\rho\rangle$, which is its eigenstate, and it acts as the identity operator on $|\phi\rangle$. Similarly,  $\cos\hat{\phi}$ and $\sin\hat{\phi}$ act nontrivially on $|\phi\rangle$, but as the identity on $|\rho\rangle$, as derived below.  Consider $\cos\hat\phi$ acting on the state $|\rho\cos\phi,\rho\sin\phi\rangle=|\rho\rangle\otimes|\phi\rangle$, which is given by
\begin{align}
    \cos\hat\phi |\rho\rangle \otimes |\phi\rangle &= \frac{\hat{r}_x}{\hat{\rho}}|\rho\rangle \otimes |\phi\rangle\\
    &= \hat{r}_x e^{- \frac{i}{\hbar} \phi \hat{L}_z} \frac{1}{\hat\rho} |\rho,0 \rangle\notag\\
    &= e^{- \frac{i}{\hbar} \phi \hat{L}_z}\underbrace{e^{ \frac{i}{\hbar} \phi\hat{L}_z}\hat{r}_x e^{- \frac{i}{\hbar} \phi \hat{L}_z}}_{\text{braiding}}\frac{1}{\hat\rho} |\rho,0 \rangle\notag\\
    &= e^{- \frac{i}{\hbar} \phi \hat{L}_z}\frac{ \cos\phi\hat{r}_x - \sin\phi\hat{r}_y}{\hat\rho} |\rho,0 \rangle\notag\\ 
     &= \cos\phi |\rho\rangle \otimes |\phi\rangle.\notag
\end{align}
Note that the eigenvalue-eigenvector relationship given above is unchanged when we change the value of $\rho$ in the radial ket.  Hence, the operator $\cos\hat\phi$ acts as the identity on the $|\rho\rangle$ ket and can be taken to act solely on the $|\phi\rangle$ ket, yielding $\cos\hat\phi|\phi\rangle = \cos\phi|\phi\rangle$. Similarly, this separation also holds for $\sin\hat\phi$. The argument for the radial operator acting only on $|\rho\rangle$ can verified by a similar argument.

We now take the limit $\delta \to 0^+$, so that $\exp\left (\frac{i}{\hbar}\vec{\delta}\cdot\hat{\vec{p}}\right) \to 1$.
Gathering the final results together, we have established that
\begin{equation} 
\label{eq:final 2D ket}
|\vec{\rho} \rangle = e^{- \frac{i}{\hbar} \rho \big(\hat{p}_\rho - i \frac{\hbar}{2\;\hat{\rho}} \big)} |\rho{=}0\rangle\otimes |\phi\rangle
\end{equation}
and
\begin{equation} 
\label{eq:final 2D bra}
\langle \vec{\rho} | =\langle \phi |\otimes\langle r{=}0|e^{  \frac{i}{\hbar}\rho \big({p}_\rho + i \frac{\hbar}{2\;\hat{\rho}} \big)}  .
\end{equation}
These final expressions are the simplification of the translation operator in plane polar coordinates that we sought when it acts on the state at the origin.

To calculate an energy wavefunction in position space, we decompose the energy eigenstate of the Hamiltonian as a tensor product of a radial and $z$-component of angular momentum state. This is given by 
\begin{equation}
|\Psi \rangle = |\psi_\rho \rangle \otimes |m \rangle ,
\end{equation}
with $\hat L_z|m\rangle=\hbar m|m\rangle$.
 The wave function is then expressed as the product of a radial wavefunction and a $z$-component of angular momentum eigenfunction according to
\begin{equation}
\langle \vec{\rho}|\Psi \rangle = \langle \rho | \psi_\rho \rangle \langle \phi | m \rangle
\end{equation}
or
\begin{equation}
\Psi (\vec{\rho}) = \psi_\rho(\rho) Y_m (\phi).
\end{equation}
Here
\begin{equation}
Y_ m (\phi) = \langle \phi|m\rangle =\langle\phi{=}0|e^{\frac{i}{\hbar}\hat L_z}|m\rangle=\langle \phi{=}0|m\rangle e^{im\phi}=  \frac{e^{im\phi}}{\sqrt{2\;\pi}}
\end{equation}
denotes the standard wavefunction of $\hat L_z$. We used the fact that normalization requires $\langle\phi{=}0|m\rangle=1/\sqrt{2\pi}$.  The radial wavefunction is then given by
\begin{equation}
\psi_\rho(r) = \langle \rho |\psi_\rho \rangle = \langle \rho{=}0 |e^{ \frac{i}{\hbar}\rho \big(\hat{p}_\rho + i \frac{\hbar}{2\;\hat{\rho}} \big)}|\psi_\rho \rangle .
\end{equation}
This is the relation that we use to determine the radial wavefunctions for the two-dimensional Hydrogen atom in the next section. 

\section{Application of the Formalism to the Coulomb Problem} 
We now show how these translation operators can be used, to calculate wavefunctions of the Coulomb problem in three and two dimensions. We start in three dimensions, where we have 
\begin{equation}
    \hat{\mathcal H}^{3d}=\frac{{\hat p}_x^2+{\hat p}_y^2+{\hat p}_z^2}{2\mu}-\frac{e^2}{\hat r}.
    \label{eq: ham_3d}
\end{equation}
Here $\mu$ is the reduced mass for the atom and $\mp e$ are the electronic and nuclear charge, respectively.
Using the decomposition of the kinetic energy into radial and angular degrees of freedom, we find that the Hamiltonian can be re-written as
\begin{equation}
    \hat{\mathcal H}^{3d}=\frac{{\hat p}_r^2}{2\mu}+\frac{{\hat{\vec L}}\cdot{\hat{\vec L}}}{2\mu\hat r^2}-\frac{e^2}{\hat r}.
    \label{eq: ham_3d_spherical}
\end{equation}
We use a tensor product to represent the statevector in terms of angular momentum eigenstates $|l,m\rangle$ and the radial state $|\psi_r\rangle$, via $|\psi\rangle=|\psi_r\rangle\otimes|l,m\rangle$. Then, because ${\hat{\vec L}}\cdot{\hat{\vec L}}|l,m\rangle=\hbar^2 l(l+1)|l,m\rangle$, we have  \begin{eqnarray}
    \hat{\mathcal H}^{3d}|\psi_r\rangle\otimes|l,m\rangle&=&\left (\frac{{\hat p}_r^2}{2\mu}+\frac{\hbar^2 l(l+1)}{2\mu\hat r^2}-\frac{e^2}{\hat r}\right )|\psi_r\rangle\otimes|l,m\rangle\nonumber\\
    &=&\hat{\mathcal H}_l^{3d}|\psi_r\rangle\otimes|l,m\rangle,
    \label{eq: tensor_product}
\end{eqnarray}
which defines the Hamiltonian $\hat{\mathcal H}_l^{3d}$ for a specific angular momentum sector. This Hamiltonian acts only on the radial part of the statevector, because all terms in $\hat{\mathcal H}_l^{3d}$ commute with $\hat{\vec{L}}$.

The Schr\"odinger factorization method has us factorize each of these angular-momentum-sector Hamiltonians via
\begin{equation} 
    \hat{\mathcal H}^{3d}_l=\frac{{\hat p}_r^2}{2\mu}+\frac{\hbar^2 l(l+1)}{2\mu\hat r^2}-\frac{e^2}{\hat r}={\hat B}_l^\dagger{\hat B}_l^{\phantom\dagger}+E_l,
    \label{eq: ham_3d_l}
\end{equation}
where we can verify that the correct lowering operator satisfies
\begin{equation}
    {\hat B}_l=\frac{1}{\sqrt{2\mu}}\left \{ {\hat p_r}-i\hbar\left ( \frac{1}{(l+1)a_0}-\frac{l+1}{\hat r}\right )\right \}
    \label{eq: 3d_Bdef}
\end{equation}
after using the commutation relation $[{\hat p}_r,\hat r^k]=-i\hbar k {\hat r}^{k-1}$ (which can be derived by induction from the radial momentum canonical commutation relation using the techniques in the appendix).  Note that one can see the reason for those specific two terms in $\hat{B}_l^{\phantom\dagger}$ is that when we square the term in parenthesis and when we commute it with $\hat{p}_r$, the resulting terms are always a constant term, a linear term in $\tfrac{1}{\hat r}$ and a quadratic term. By adjusting the coefficients, we can arrange for the product ${\hat B}^\dagger_l{\hat B}^{\phantom\dagger}_l$ to have the required form. The symbol $a_0=\tfrac{\hbar^2}{\mu e^2}$ is the Bohr radius and $E_l=-\tfrac{e^2}{2(l+1)^2a_0}$.

%Note that because $\hat B}_l^\dagger{\hat B}_l^{\phantom\dagger}$ is a positive semidefinite operator, %its ground state satisfies
%\begin{equation}
%    {\hat B}_l^{\phantom\dagger}|n{=}l+1,l\ran
%with energy $E_l$. We used the symbolgle=0,
%    \label{eq: 3d_subsidiary}
%\end{equation} $n$ to denote the principal quantum number, which here is equal to %$l+1$. In terms of the principal quantum number, we have the energy of the state $|n,l{=}n-1\rangle$ is %$-\tfrac{e^2}{2n^2a_0}$. It is unfortunate that within this notation, the energy for the state %$|n,n-1\rangle$ is given by $E_{n-1}$.

We denote the eigenstate of $\hat{\mathcal H}^{3d}_l$, corresponding to eigenvalue $E_{l{=}n-1}$, as $|n,l{=}n-1\rangle$.  Here $n$, which is a positive integer, is the standard principal quantum number, and the energy is degenerate for all $l$ such that $0\le l\le n-1$.  We then have
\begin{equation}
E_{n-1}=-\frac{e^2}{2\;n^2a_0} \;, 
\end{equation}
which, unfortunately, is conventionally denoted as $E_n$. We don't redefine it, as is customarily done, to avoid confusion in the formulas that follow. Note further that the choice we made for the lowering operator in Eq.~(\ref{eq: 3d_Bdef}) was made so that the set of energies $E_{n-1}$ form an \textit{increasing sequence} for $1\le n< \infty$. This choice for the increasing sequence guarantees that the eigenfuntions are all normalizable bound states.

Since $\hat{B}_l^{\dagger}\hat{B}_l^{\phantom{\dagger}}$ is a non-negative semidefinite operator, it follows from Eq. \eqref{eq:  ham_3d_l} that the ground state $|n,n-1\rangle$ of $\hat{\mathcal H}^{3d}_{n-1}$ satisfies
\begin{equation}
 \hat{\mathcal H}^{3d}_{n-1}|n,n-1\rangle = E_{n-1} |n,n-1\rangle \; ,
\end{equation}
with
\begin{equation} \label{eq:3d_subsidiary}
 \hat{B}_{n-1} |n,n-1\rangle = 0 .  
\end{equation}
As we will see that this condition, which we term the \emph{subsidiary condition}, facilitates the determination of all of the eigenstates $|n,l\rangle$.

One can also directly verify that when the raising and lowering operators act in the ``wrong'' order, we have
\begin{equation}
    {\hat B}_l^{\phantom\dagger}{\hat B}_l^{\dagger}=\frac{{\hat p}_r^2}{2\mu}+\frac{\hbar^2 (l+1)(l+2)}{2\mu\hat r^2}-\frac{e^2}{\hat r}-E_l=\hat{\mathcal H}_{l+1}^{3d}-E_l.
    \label{eq: 3d_intertwining}
\end{equation}
This allows us to derive the intertwining relationship, when a ${\hat B}^\dagger_l$ operator is moved to the left past a $\hat{\mathcal H}^{3d}_l$ operator. In particular, we have
\begin{eqnarray}
\hat{\mathcal H}^{3d}_l{\hat B}^\dagger_l&=&\left ({\hat B}_l^\dagger{\hat B}_l^{\phantom\dagger}+E_l\right ){\hat B}_l^\dagger\nonumber\\
&=&{\hat B}_l^\dagger{\hat B}_l^{\phantom\dagger}{\hat B}_l^\dagger+E_l{\hat B}_l^\dagger\nonumber\\
&=&{\hat B}_l^\dagger\left ({\hat B}_l^{\phantom\dagger}{\hat B}_l^\dagger+E_l\right )\nonumber\\
&=&{\hat B}_l^\dagger\hat{\mathcal H}^{3d}_{l+1}.
\label{eq: 3d_intertwining2}
\end{eqnarray}
Hence when a ${\hat B}_l^\dagger$ operator starts on the right and moves to the left through an $\hat{\mathcal H}_l^{3d}$, it shifts the index of the Hamiltonian upward by one unit.

We use this intertwining identity to construct all of the energy eigenstates that have energy $E_{n-1}$.   Note that this approach is different from the original Schr\"odinger approach. It is the simplest way to establish all of the energy eigenstates.  For each $n$, we can find a total of $n$ states with different total angular momentum that are eigenstates. The key observation is that the following set of states are eigenstates:
\begin{equation}
|n,l\rangle=C_{nl}{\hat B}_l^\dagger{\hat B}_{l+1}^\dagger\cdots {\hat B}_{n-3}^\dagger{\hat B}_{n-2}^\dagger|n,n-1\rangle ,
\label{eq: 3d_eigenstate}
\end{equation}
where $0\le l\le n-1$: note that when $l=n-1$, there are \textit{no} ${\hat B}^\dagger$ operators needed and when $l=0$, the string runs from $l=n-2$ down to $l=0$. In particular, there is \textit{no} ${\hat B}_{-1}^\dagger$ operator, because $E_l \to - \infty$ as $l \to - 1$.  This terminates the chain at $l=0$. The number $C_{nl}$ is a normalization constant, which we determine below. To prove that $|n,l\rangle$ is an eigenstate of energy $E_{n-1}$, we simply operate with $\hat{\mathcal H}_l^{3d}$ on $|n,l\rangle$. Using the intertwining relation, to move the $\mathcal{H}^{3d}$ operator to the right, increasing its index by one with every step, we find that
\begin{eqnarray}
    \hat{\mathcal H}_l^{3d}|n,l\rangle&=&\hat{\mathcal H}_l^{3d}C_{nl}{\hat B}_l^\dagger{\hat B}_{l+1}^\dagger\cdots {\hat B}_{n-3}^\dagger{\hat B}_{n-2}^\dagger|n,n-1\rangle\nonumber\\
    &=&C_{nl}{\hat B}_l^\dagger\hat{\mathcal H}_{l+1}^{3d}{\hat B}_{l+1}^\dagger\cdots {\hat B}_{n-3}^\dagger{\hat B}_{n-2}^\dagger|n,n-1\rangle\nonumber\\
    &=&C_{nl}{\hat B}_l^\dagger{\hat B}_{l+1}^\dagger\cdots {\hat B}_{n-3}^\dagger\hat{\mathcal H}_{n-2}^{3d}{\hat B}_{n-2}^\dagger|n,n-1\rangle\nonumber\\
    &=&C_{nl}{\hat B}_l^\dagger{\hat B}_{l+1}^\dagger\cdots {\hat B}_{n-3}^\dagger{\hat B}_{n-2}^\dagger\hat{\mathcal H}_{n-1}^{3d}|n,n-1\rangle\nonumber\\
    &=&-\frac{e^2}{2n^2a_0}|n,l\rangle,
    \label{eq: 3d_verify_eigenstate}
\end{eqnarray}
since the state $|n,n-1\rangle$ is an eigenstate of $\hat{\mathcal H}_{n-1}^{3d}$ with eigenvalue $-e^2/2n^2a_0$.  We see that for a given $n$, all of the states $|n,l\rangle$, with $0 \le l \le n - 1$, belong to the eigenvalue $E_{n - 1}$.  So we have constructed eigenstates of each of the $l$th angular-momentum-sector Hamiltonians with $l\le n-1$. When we make a tensor product of such a state with a $|l,m\rangle$ angular momentum state, we obtain an eigenstate of the full Coulomb Hamiltonian.

\begin{figure}[H]
\centerline{\includegraphics[width=3.3in]{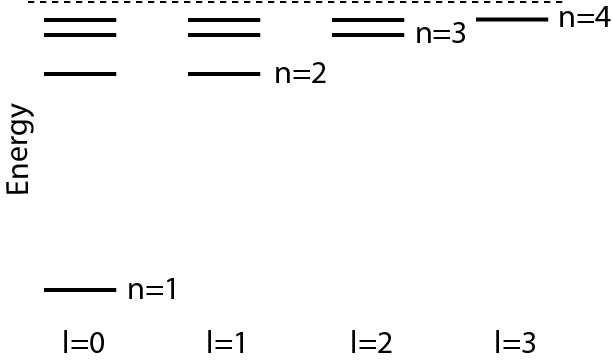}}
\caption{Plot of the energy levels of the three-dimensional Coulomb problem up to $n=4$. The states $|n,n-1\rangle$  are the rightmost states in each row. Each additional $\hat B^\dagger$  operator moves us one step to the left (there are no more $\hat B^\dagger$ operators to apply after $l=0$). All energies in a given row are degenerate. The dashed line shows where $E=0$.}
\label{fig: 3d_energy_schematic}
\end{figure}

Before moving further in the derivation, we illustrate schematically what the energy levels are and how the different eigenstates interrelate. This is depicted in Fig.~\ref{fig: 3d_energy_schematic}.

We also use the intertwining relation to normalize the state and determine $C_{nl}$. This is done by simply computing the norm
\begin{equation}
    1=\langle n,l|n,l\rangle=|C_{nl}|^2\langle n,n-1|{\hat B}_{n-2}^{\phantom\dagger}\cdots \underbrace{{\hat B}_l^{\phantom\dagger}{\hat B}_l^\dagger}\cdots {\hat B}_{n-2}^\dagger|n,n-1\rangle.
    \label{eq: 3d_norm}
\end{equation}
Then we start with the innermost pair of raising and lowering operators and note that ${\hat B}_l^{\phantom\dagger}{\hat B}_l^\dagger=\hat{\mathcal H}_{l+1}^{3d}-E_l$. If we move this operator through to the right, increasing its index by one with each step,  until it acts directly on the state $|n,n-1\rangle$, we see that the intertwining relation will convert the Hamiltonian to the one corresponding to $l=n-1$. This can be immediately evaluated against the state $|n,n-1\rangle$ yielding the eigenvalue $E_{n-1}$. The net effect is we remove the product of the two operators ${\hat B}_l^{\phantom\dagger}{\hat B}_l^\dagger$ and replace it with the factor $(E_{n-1}-E_l)$. Repeating $n-l-2$ more times, we find that
\begin{eqnarray}
\label{eq: 3d_norm1}
    C_{nl}&=&\frac{1}{\sqrt{\prod_{k=l}^{n-2}(E_{n-1}-E_k)}}\\
    &=&\sqrt{\left (\frac{2a_0n^2}{e^2}\right )^{n-l-1}\frac{(n+l)![(n-1)!]^2}{(2n-1)!(n-l-1)!(l!)^2}} .\label{eq: 3d_norm2}
\end{eqnarray} 
Note that this final result requires that the initial state $|n,n-1\rangle$ is normalized, i.e., $\langle n,n-1|n,n-1 \rangle = 1$.  We will ensure that this is true.

The next step is to show that the string of $\hat{B}^{\dagger}$ operators acting on $|n,n-1 \rangle$ is proportional to a Laguerre polynomial in $\hat r$ (multiplied by a power of $1/\hat{r}$ (acting on the same state. To do this, we need to go through a proof by induction. Our starting point  to observe that the subsidiary condition in Eq.~\eqref{eq:3d_subsidiary} can be rewritten as
\begin{equation}
    {\hat p}_r|n,n-1\rangle=i\hbar\left (\frac{1}{na_0}-\frac{n}{\hat r}\right )|n,n-1\rangle.
    \label{eq: 3d_subsidiary2}
\end{equation}
This allows us to replace the radial momentum operator acting on the state $|n,n-1\rangle$ by the sum of a constant and a term that goes like $1/\hat r$ (acting on the same state). Given that the commutation relation of the radial momentum with an inverse power of $\hat r$ increases the power by one as well, one can immediately see that the string of ${\hat B}^\dagger$ operators acting on $|n,n-1\rangle$ will be a polynomial of degree $n-l-1$ in $1/\hat r$.  By factoring out $1/{\hat r}^{n-l-1}$, we are left with a polynomial of degree $n-l-1$ in $\hat r$. Recognizing that the dimensionality of a ${\hat B}^\dagger$ operator is $\hbar/(\sqrt{\mu}a_0)$ and using $2{\hat r}/na_0$ as the expansion parameter for the polynomial, we have that
\begin{eqnarray}
    {\hat B}_l^\dagger\cdots {\hat B}_{n-2}^\dagger|n,n-1\rangle&=&
   \left ( \frac{2i\hbar}{\sqrt{2\mu}na_0}\right )^{n-l-1}\left (\frac{n a_0}{2\hat r}\right )^{n-l-1}\nonumber\\
   &\times&\sum_{j=0}^{n-l-1}b^{(l)}_j\left (\frac{2\hat r}{na_0}\right )^{j}|n,n-1\rangle.
    \label{eq: 3d_polynomial_def}
\end{eqnarray}
Note that we are suppressing a label of $n$ corresponding to the principal quantum number in the notation for the $b$ coefficients to streamline the notation; one must remember that a given set of $b$ coefficients is generated for each principal quantum number $n$.
We choose the base case to corresponds to $l=n-2$, where the polynomial is just equal to 
\begin{eqnarray}
    &~&\hat B^\dagger_{n-2}|n,n-1\rangle=\frac{i\hbar}{\sqrt{2\mu}}\left [ \frac{2n-1}{n(n-1)a_0}-\frac{2n-1}{\hat r}\right ]|n,n-1\rangle\nonumber\\
    &~&~~=\left ( \frac{2i\hbar}{\sqrt{2\mu}na_0}\right )\left (\frac{n a_0}{2\hat r}\right )\left [ -2n+1+\frac{2n-1}{2(n-1)}\left (\frac{2\hat r}{na_0}\right )\right ],\nonumber\\
    &~&~~~~~~~~~~~\times|n,n-1\rangle\label{eq: base_poly_3d}
\end{eqnarray}
so that $b_0^{(n-2)}=-2n+1$ and $b_1^{(n-2)}=(2n-1)/2(n-1)$ (one could have taken the base case as $l=n-1$, with $b_0^{(n-1)}=1$, but we chose not to because that is a trivial base case).
To evaluate the induction step, we remove the leftmost ${\hat B}^\dagger_l$ and collect what remains in terms of the polynomial for $l+1$. Then we operate the ${\hat p}_r$ operator to the right, commuting it through the $\hat r$ operators until it can act on the state $|n,n-1\rangle$, where we can use Eq.~(\ref{eq: 3d_subsidiary2}) to replace it by a monomial in $1/\hat r$. After collecting the coefficients of powers of $\hat r$, one finds
\begin{eqnarray}
    \sum_{j=0}^{n-l-1}b_j^{(l)}\left (\frac{2\hat r}{na_0}\right )^j&=&\sum_{j=0}^{n-l-1}\left (\frac{2\hat r}{na_0}\right )^{j}\label{eq: induction_3d}\\
    &\times&\left [ -(j+2l+3)b_j^{(l+1)}+\frac{n+l+1}{2(l+1)}b_{j-1}^{(l+1)}\right ],\nonumber
\end{eqnarray}
where we set $b_{-1}^{(l+1)}=b_{n-l-1}^{(l+1)}=0$. The key to determining that these results yield associated Laguerre polynomials is to compute the ratio of successive coefficients of the $l$th polynomial using the two formulas we have. This gives
\begin{equation}
    \frac{b_{j+1}^{(l)}}{b_j^{(l)}}=
    \frac{{\scriptstyle(j+2l+4)}\left (\frac{b_{j+1}^{(l+1)}}{b_j^{(l+1)}}\right ){\scriptstyle -}\frac{n+l+1}{2(l+1)}}{{\scriptstyle(j+2l+3)-}\frac{n+l+1}{2(l+1)}\left (\frac{b_j^{(l+1)}}{b_{j-1}^{(l+1)}}\right )^{-1}}~~\text{for}~~0\le j\le n-l-2.
    \label{eq: ratio_3d}
\end{equation}
Here again, we note that $b_{-1}^{(l+1)}=b_{n-l-1}^{(l+1)}=0$. This brings us to the induction hypothesis. It is that the ratio is given by
\begin{equation}
    \frac{b_{j+1}^{(l)}}{b_j^{(l)}}=\frac{j-n+l+1}{(j+1)(j+2l+2)}.
    \label{eq: induction_hypothesis_3d}
\end{equation}
One can immediately see from Eq.~(\ref{eq: base_poly_3d}) for $l=n-2$, that the base case gives $b_1^{(n-2)}/b_0^{(n-2)}=-1/2(n-1)$, which agrees with the induction hypothesis in Eq.~(\ref{eq: induction_hypothesis_3d}) for $l=n-2$ and $j=0$. So, we assume it holds for $l+1$ and substitute into Eq.~(\ref{eq: ratio_3d}) to show it holds for $l$. This yields
\begin{equation}
    \frac{b_{j+1}^{(l)}}{b_j^{(l)}}=
    \frac{\frac{(j-n+l+2)}{(j+1)}-\frac{(n+l+1)}{2(l+1)}}{{\scriptstyle(j+2l+3)}\left [1-\frac{(n+l+1)j}{2(l+1)(j-n+l+1)}\right ]}.
\end{equation}
Putting everything over common denominators yields
\begin{equation}
    \frac{b_{j+1}^{(l)}}{b_j^{(l)}}=
    \tfrac{(j-n+l+1)[2(l+1)(j-n+l+2)-(j+1)(n+l+1)]}{(j+2l+3)(j+1)[2(l+1)(j-n+l+1)-j(n+l+1)]}.
\end{equation}
Expanding the terms in the numerator and denominator in the square brackets, factorizing the expansion, and simplifying, then reduces this result to the one in Eq.~(\ref{eq: induction_hypothesis_3d}), which establishes the induction proof. It turns out that the coefficients of a polynomial that satisfy Eq.~(\ref{eq: induction_hypothesis_3d}) are associated Laguerre functions. We use the definition of the associated Laguerre function given by Powell and Crasemann in their quantum mechanics textbook~\cite{powell_crasemann}:
\begin{equation}
    L_m^{(\alpha)}=\sum_{j=0}^{m}\frac{(-1)^j}{j!}\left ( \begin{array}{c}
    m+\alpha\\m-j\end{array}\right )\,x^j=\sum_{j=0}^m a_j^{(\alpha,m)}x^j,
    \label{eq: laguerre_def}
\end{equation}
which defines the coefficients of the polynomial $a_j^{(\alpha,m)}$. We immediately see from Eq.~(\ref{eq: laguerre_def}) that
\begin{equation}
    \frac{a_{j+1}^{(\alpha,m)}}{a_j^{(\alpha,m)}}=\frac{j-m}{(j+1)(\alpha+j+1)}.
    \label{eq: laguerre_ratio}
\end{equation}
This then tells us that the polynomial given by the string of $\hat B^\dagger$ operators acting on $|n,n-1\rangle$ is {\it proportional} to the associated Laguerre polynomial with argument given by $2\hat r /na_0$, $\alpha=2l+1$ and $m=n-l-1$. To find the constant of proportionality [in addition to the power $(2\hat r /na_0)^{n-l-1}$, which we already knew from Eq.~(\ref{eq: 3d_polynomial_def})], we evaluate the constant term (no $\hat r$ dependence or the maximal power of $\hat r$ in the sum) in Eq.~(\ref{eq: 3d_polynomial_def}). It is given by
\begin{eqnarray}
    &~&\left (\frac{2i\hbar}{\sqrt{2\mu}na_0}\right )^{n-l-1}b_{n-l-1}^{(l)}\nonumber\\
    &~&~~~~~=\left ( \frac{i\hbar}{\sqrt{2\mu}}\right )^{n-l-1}\prod_{k=l+1}^{n-1}\left (\frac{1}{na_0}+\frac{1}{ka_0}\right ),
    \label{eq: constant_term}
\end{eqnarray}
where the right hand side is found from the constant terms in each $\hat B^\dagger$ and from each ${\hat p}_r$ in each $\hat B^\dagger$ when it acts on $|n,n-1\rangle$. This constant term is not modified by the commutators of the radial momentum with inverse powers of $\hat r$. From this, we learn that
\begin{equation}
    b_{n-l+1}^{(l)}=\left (\frac{1}{2}\right )^{n-l-1}\frac{l!(2n-1)!}{(n-1)!(n+l)!}.
    \label{eq: b_value}
\end{equation}
But if we instead express in terms of the maximal coefficient of the Laguerre polynomial, we find that
\begin{equation}
     b_{n-l+1}^{(l)}=C'a_{n-l-1}^{(2l+1,n-l-1)}=C'\frac{(-1)^{n-l-1}}{(n-l-1)!},
     \label{eq: a_in_terms_of_b}
\end{equation}
where $C'$ is the proportionality constant we need to determine. We immediately learn that
\begin{equation}
    C'=\left ( -\frac{1}{2}\right )^{n-l-1}\frac{l!(n-l-1)!(2n-1)!}{(n+l)!(n-1)!}.
\end{equation}
Putting this all together, we have derived the identity that
\begin{eqnarray}
&~& {\hat B}_l^\dagger{\hat B}_{l+1}^\dagger\cdots {\hat B}_{n-3}^\dagger{\hat B}_{n-2}^\dagger|n,n-1\rangle\nonumber\\
    &~&~~~~~~=\left ( -\frac{i\hbar}{\sqrt{2\mu}na_0}\right )^{n-l-1}\frac{l!(n-l-1)!(2n-1)!}{(n+l)!(n-1)!}\nonumber\\
   &~&~~~~~~~~~\times
   \left (\frac{n a_0}{2\hat r}\right )^{n-l-1}L_{n-l-1}^{2l+1}\left (\frac{2\hat r}{na_0}\right )|n,n-1\rangle.
\end{eqnarray}
Finally, we multiply by $C_{nl}$ from Eq.~(\ref{eq: 3d_norm2}) and find that
\begin{eqnarray}
|n,l\rangle&=&(-i)^{n-l-1}\sqrt{\frac{(n-l-1)!(2n-1)!}{(n+l)!}}\left (\frac{n a_0}{2\hat r}\right )^{n-l-1}\nonumber\\
&~&~~~~~\times L_{n-l-1}^{2l+1}\left (\frac{2\hat r}{na_0}\right )|n,n-1\rangle.
\end{eqnarray}
Since the eigenvector is defined only up to a complex phase, we choose to remove the complex phase in the following. 

We are now ready to compute the wavefunction. We take the overlap of the position eigenstate with the energy eigenstate, or
\begin{equation}
    (\langle\theta\phi|\otimes\langle r|)(|n,l\rangle\otimes|l,m\rangle).
    \label{eq: 3d_wavefunction}
\end{equation}
Using the fact that~\cite{weitzman_freericks}
\begin{equation}
Y_{lm}=\langle\theta\phi|lm\rangle,
\label{eq: spherical_harmonic}
\end{equation}
we find that
\begin{eqnarray}
&~&\psi_{nlm}(r,\theta,\phi)=\sqrt{\frac{(n-l-1)!(2n-1)!}{(n+l)!}}\\
&~&\times \langle r|\left (\frac{n a_0}{2\hat r}\right )^{n-l-1}L_{n-l-1}^{2l+1}\left (\frac{2\hat r}{na_0}\right )|n,n-1\rangle Y_{lm}(\theta,\phi).\nonumber
\label{eq: 3d_wf1}
\end{eqnarray}
Because $\hat r|r\rangle=r|r\rangle$, we immediately find that
\begin{eqnarray}
&~&\psi_{nlm}(r,\theta,\phi)=\sqrt{\frac{(n-l-1)!(2n-1)!}{(n+l)!}}\langle r|n,n-1\rangle\nonumber\\
&~&~~~~~~~~~~\times \left (\frac{n a_0}{2 r}\right )^{n-l-1}L_{n-l-1}^{2l+1}\left (\frac{2 r}{na_0}\right )Y_{lm}(\theta,\phi).
\label{eq: 3d_wf2}
\end{eqnarray}
What remains is to determine the first wavefunction in the chain, $\phi_n(r)=\langle r|n,n-1\rangle$. This is where we need to use the translation operator in spherical coordinates. The radial translation operator is an exponential of $ \left ( {\hat p}_r+\tfrac{i\hbar}{\hat r}\right )$, when acting on the position eigenvector at the origin in position space. While one can evaluate the exponential of an operator acting on a state via expanding the power series term by term, a more efficient evaluation can be accomplished if one evaluates the operator acting on one of its eigenvectors---then the exponentiation becomes trivial. This is the strategy we adopt here. 

We start with some additional operator identities. First note that $[{\hat p}_r,\hat r^{-n+1}]=i\hbar (n-1)/\hat r^{n}$ (see the Appendix for details). Hence, we use Eq.~(\ref{eq: 3d_subsidiary2}) and the commutator to find that
\begin{equation}
    {\hat p}_r\frac{1}{\hat r^{n-1}}|n,n-1\rangle=i\hbar \left (\frac{1}{na_0\hat r^{n-1}}-\frac{1}{\hat r^n}\right )|n,n-1\rangle,
    \label{eq: rad_mom_on_phi}
\end{equation}
or
\begin{equation}
    \left ( {\hat p}_r+\frac{i\hbar}{\hat r}\right )\frac{1}{\hat r^{n-1}}|n,n-1\rangle
    =\frac{i\hbar}{na_0}\frac{1}{\hat r^{n-1}}|n,n-1\rangle.
    \label{eq: rad_mom_on_phi2}
\end{equation}
Note that this says that the state $(1/\hat r^{n-1})|n,n-1\rangle$ is an eigenvector for the operator $ \left ( {\hat p}_r+\tfrac{i\hbar}{\hat r}\right )$ with eigenvalue $i\hbar/na_0$. An imaginary eigenvalue is possible because this operator is \textit{not} Hermitian. 

We are now ready to use the translation operator in spherical coordinates. Since $\langle \theta\phi|=\langle 0,0,1|\exp(i\theta{\hat L}_y/\hbar)\exp(i\phi{\hat L}_z/\hbar)$, we have that 
\begin{equation}
    \langle r|=\langle 0_r|\exp\left [ \frac{ir}{\hbar}\left ( {\hat p}_r+\frac{i\hbar}{\hat r}\right )\right ].
\end{equation}
The wavefunction $\phi_n(r)$ then becomes
\begin{eqnarray}
    \phi_n(r)&=&\langle r|n,n-1\rangle=\langle r|\frac{\hat r^{n-1}}{\hat r^{n-1}}|n,n-1\rangle=r^{n-1}\langle r|\frac{1}{\hat r^{n-1}}|n,n-1\rangle\nonumber\\
    &=&r^{n-1}\langle 0_r|\exp\left [ \frac{ir}{\hbar}\left ( {\hat p}_r+\frac{i\hbar}{\hat r}\right )\right ]\frac{1}{\hat r^{n-1}}|n,n-1\rangle.
\end{eqnarray}
Because the state on the right is an eigenvector for $ \left ( {\hat p}_r+\tfrac{i\hbar}{\hat r}\right )$, we immediately find that
\begin{equation}
     \phi_n(r)=r^{n-1}\exp\left (-\frac{r}{na_0}\right )\langle 0_r|\frac{1}{\hat r^{n-1}}|n,n-1\rangle.
     \label{eq: 3d_phi_final}
\end{equation}
Note that the term $\langle 0_r|\frac{1}{\hat r^{n-1}}|n,n-1\rangle$ is a constant because Eq.~(\ref{eq: 3d_phi_final})
shows that
\begin{equation}
    \lim_{r\to 0}\frac{1}{r^{n-1}}\phi_n(r)=\langle 0_r|\frac{1}{\hat r^{n-1}}|n,n-1\rangle;
\end{equation}
its precise value is determined by normalization. We find
\begin{equation}
    1=\left (\langle 0_r|\frac{1}{\hat r^{n-1}}|n,n-1\rangle\right )^2\int_0^\infty \,dr\, r^{2n}e^{-\frac{2r}{na_0}},
    \label{eq: 3d_phi_norm}
\end{equation}
which yields
\begin{equation}
    \phi_n(r)=\left (\frac{2}{na_0}\right )^{n+\frac{1}{2}}\frac{1}{\sqrt{(2n)!}} r^{n-1}\exp\left (-\frac{r}{na_0}\right ).
    \label{eq: 3d_phi_final2}
\end{equation}
While this result can be calculated in many different ways, one cannot calculate this wavefunction using the translation operator in Cartesian coordinates; it must be in spherical coordinates, because it is a power series in $r$ that includes odd powers.

We can now summarize our final wavefunction by combining all of our results together. It is
\begin{eqnarray}
\psi_{nlm}(r,\theta,\phi)&=&\left (\frac{2}{na_0}\right )^{\frac{3}{2}}\sqrt{\frac{(n-l-1)!}{2n(n+l)!}}\\
&\times& \left (\frac{2 r}{na_0}\right )^{l}L_{n-l-1}^{2l+1}\left (\frac{2 r}{na_0}\right )e^{-\frac{r}{na_0}}Y_{lm}(\theta,\phi),\nonumber
\label{eq: 3d_wf_final}
\end{eqnarray}
which is the standard result for the three-dimensional Coulomb bound-state wavefunctions using the Laguerre polynomial definition in Eq.~(\ref{eq: laguerre_def}). We want to stress that the entire calculation was based on the representation-independent approach described in the introduction. All of the steps in the derivation used operator algebra. We never needed to represent momentum operators in terms of spatial derivatives. It is comforting to know that wavefunctions \textit{can} be calculated in such a representation independent way.

Having completed our first example, we now move on to the second. We will expedite the description, because many of the techniques for the two-dimensional case are similar to those of the three-dimensional case. But the two-dimensional case does illustrate some interesting new twists, so it is important to describe it carefully.

The Coulomb Hamiltonian in two dimensions is given by
\begin{equation}
    \hat{\mathcal H}^{2d}=\frac{{\hat p}_x^2+{\hat p}_y^2}{2\mu}-\frac{e^2}{\hat \rho}.
    \label{eq: ham_2d}
\end{equation}
The kinetic energy can again be decomposed into radial and angular components (see the Appendix), 
\begin{equation}
    \mathcal{H}^{2d}=\frac{{\hat p}_\rho^2}{2\mu}+\frac{{\hat L}_z^2-\tfrac{\hbar^2}{4}}{2\mu\hat\rho^2}-\frac{e^2}{\hat\rho},
    \label{eq: ham_2d2}
\end{equation}
with ${\hat p}_{\rho}={\hat r}_x{\hat p_x}+{\hat r_y}{\hat p_y}-\tfrac{i\hbar}{2\hat\rho}$. Note that in two dimensions we do have a quantum correction to the kinetic energy (given by the $-\hbar^2/8\mu\hat\rho^2$ term). Here, the angular-momentum states are given by eigenstates of ${\hat L}_z$, which satisfy ${\hat L}_z|m\rangle=\hbar m|m\rangle$. So, we form the energy eigenstates in terms of a tensor product of radial and angular momentum states via $|\psi\rangle=|\psi_\rho\rangle\otimes|m\rangle$. Operating the Hamiltonian onto the tensor-product state yields
\begin{eqnarray}
    \mathcal{H}^{2d}|\psi_\rho\rangle\otimes|m\rangle&=&\left (\frac{{\hat p}_\rho^2}{2\mu}+\frac{\hbar^2(m^2-\tfrac{1}{4})}{2\mu\hat\rho^2}-\frac{e^2}{\hat\rho}\right )|\psi_\rho\rangle\otimes|m\rangle\nonumber\\
    &=&\mathcal{H}^{2d}_m|\psi_\rho\rangle\otimes|m\rangle,
    \label{eq: 2d_ham_m}
\end{eqnarray}
where the second line defines the set of Hamiltonians $\{\mathcal{H}_m^{2d}: m\in\mathbb{Z}\}$ that operate only on the radial state $|\psi_\rho\rangle$. Note that here, the index $m$ can be positive or negative, unlike in the three-dimensional case, where $l$ is a nonnegative integer. Since each of these Hamiltonians depends on $m$ only through $m^2$, we immediately learn that the radial eigenfunctions and the energy eigenvalues depend only on the magnitude of $m$, that is on $|m|$. 

We continue to use the Schr\"odinger factorization method for $m\ge 0$ via
\begin{equation}
    \mathcal{H}^{2d}_m=\frac{{\hat p}_{\rho}^2}{2\mu}+\frac{\hbar^2\left (m^2-\tfrac{1}{4}\right )}{2\mu\hat\rho^2}-\frac{e^2}{\hat\rho}={\hat B}_m^\dagger{\hat B}_m^{\phantom\dagger}+E_m.
    \label{eq: 2d_factorization}
\end{equation}
A quick calculation tells us that
\begin{equation}
    {\hat B}_m^{\phantom\dagger}=\frac{1}{\sqrt{2\mu}}\left [ {\hat p}_\rho-i\hbar\left (\frac{1}{\left (m+\tfrac{1}{2}\right )a_0}-\frac{\left (m+\tfrac{1}{2}\right )}{\hat\rho}\right )\right ]
    \label{eq: 2d_B_def}
\end{equation}
and
\begin{equation}
    E_m=-\frac{e^2}{2a_0\left (m+\tfrac{1}{2}\right )^2}.
    \label{eq: 2d_energy}
\end{equation}
Note that for $m\ge 0$, the energies form an increasing sequence, which is required for the energy eigenstates to all be normalizable.
While the algebra above holds for all $m$, we first focus on working with nonnegative $m$; we will describe how to handle negative $m$ later. In particular, we already know that the radial part of the eigenvectors will be identical, as will the energies, so we will not just be using the above formulas with negative $m$ to solve these problems.

Computing the product of the two operators in opposite order gives us
\begin{equation}
    {\hat B}_m^{\phantom\dagger}{\hat B}_m^\dagger+E_m=\frac{{\hat p}_{\rho}^2}{2\mu}+\frac{\hbar^2\left ((m+1)^2-\tfrac{1}{4}\right )}{2\mu\hat\rho^2}-\frac{e^2}{\hat\rho}=\mathcal{H}_{m+1}^{2d}.
    \label{eq: 2d_factorization2}
\end{equation}
This relation is the same relation as we had in three-dimensions, so, we immediately find the corresponding intertwining relation (for $m\ge 0$)
\begin{equation}
    \mathcal{H}_{m}^{2d}{\hat B}_m^\dagger={\hat B}_m^\dagger\mathcal{H}_{m+1}^{2d},
    \label{eq: 2d_intertwining}
\end{equation}
which can be employed to find eigenstates in the same fashion: we have the eigenvectors given by
\begin{equation}
     |n,m\rangle=C_{nm}{\hat B}_m^\dagger{\hat B}_{m+1}^\dagger\cdots{\hat B}_{n-3}^\dagger{\hat B}_{n-2}^\dagger|n,n-1\rangle
    \label{eq: 2d_eigenvector}
\end{equation}
for $m\ge 0$, with the eigenvalues equal to
\begin{equation}
    E_{n-1}=-\frac{e^2}{2a_0\left (n-\tfrac{1}{2}\right )^2}
    \label{eq: 2d_energy2}
\end{equation}
and $C_{nm}$ a normalization constant.
The state $|n,m{=}n-1\rangle=|n,n-1\rangle$ satisfies 
\begin{equation}
    {\hat B}_{n-1}^{\phantom\dagger}|n,n-1\rangle=0.
    \label{eq: 2d_subsidiary}
\end{equation}
As we saw in the three-dimensional case, the normalization constant (for $m\ge 0$) is given by
\begin{eqnarray}
    C_{nm}&=&\frac{1}{\sqrt{\prod_{k=m}^{n-2}(E_{n-1}-E_k)}}\nonumber\\
    &=&\left (\frac{\sqrt{a_0}\left (n-\tfrac{1}{2}\right )}{\sqrt{2}e}\right )^{n-m-1}\nonumber\\
    &\times&\sqrt{\frac{(n+m-1)![(2n-3)!!]^2}{(2n-2)!(n-m-1)![(2m-1)!!]^2}}.
    \label{eq: 2d_cmn}
\end{eqnarray}

We now discuss some interesting observations about what happens when $m<0$. For example, one can see from Eq.~(\ref{eq: 2d_B_def}) that ${\hat B}_{-|m|}^{\phantom\dagger}={\hat B}_{|m|-1}^\dagger$. So, if we extend the eigenstates in Eq.~(\ref{eq: 2d_eigenvector}) for $m<0$, then, because we have that
\begin{eqnarray}
    &~&{\hat B}^\dagger_{-|m|}\cdots {\hat B}_{-1}^\dagger{\hat B}_0^\dagger\cdots{\hat B}_{|m|-1}^\dagger\nonumber\\
    &~&~~~~~~~={\hat B}^{\phantom\dagger}_{|m|-1}\cdots \underbrace{{\hat B}_{0}^{\phantom\dagger}{\hat B}_0^\dagger}\cdots{\hat B}_{|m|-1}^\dagger\nonumber\\
    &~&~~~~~~~={\hat B}^{\phantom\dagger}_{|m|-1}\cdots{\hat B}_1^{\phantom\dagger} (\mathcal{H}_1^{2d}-E_0){\hat B}_1^\dagger\cdots{\hat B}_{|m|-1}^\dagger\nonumber\\
    &~&~~~~~~~={\hat B}^{\phantom\dagger}_{|m|-1}\cdots {\hat B}_{1}^{\phantom\dagger}{\hat B}_1^\dagger\cdots{\hat B}_{|m|-1}^\dagger(\mathcal{H}_{|m|}^{2d}-E_0)\nonumber\\
    &~&~~~~~~~=(\mathcal{H}_{|m|}^{2d}-E_{|m|-1})\cdots(\mathcal{H}_{|m|}^{2d}-E_0),
    \label{eq: minus_m_identity}
\end{eqnarray}
we can show the equivalence of the radial component of the eigenvectors for negative $m$ with positive $m$. 
If we let the product of operators in Eq.~(\ref{eq: minus_m_identity}) act on $|n,|m|\rangle$, then each $\mathcal{H}_{|m|}^{2d}$ yields $E_{|m|}$, so we find that
\begin{equation}
    |n,-|m|\rangle=|n,|m|\rangle
    \label{eq: minus_m_state_identity}
\end{equation}
directly from the operator identities. This confirms our statement before that the radial component of the eigenvector for $-|m|$ is \textit{identical} to the eigenvector for $|m|$ (with the same principal quantum number $n$). Note further that the product of numerical factors are precisely the numerical factors to guarantee that $C_{n-|m|}=C_{n|m|}$, which is why we have an equality in Eq.~(\ref{eq: minus_m_state_identity}) rather than a proportionality.

\begin{figure}[htb]
\centerline{\includegraphics[width=3.3in]{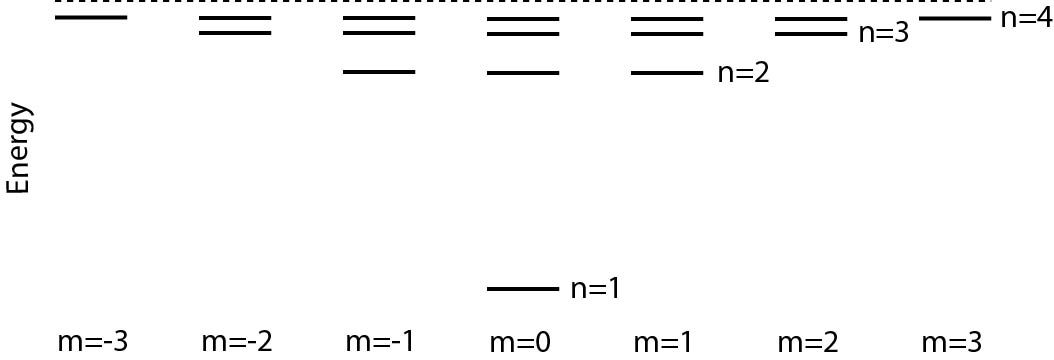}}
\caption{Plot of the energy levels of the two-dimensional Coulomb problem up to $n=4$. The states $|n,n-1\rangle$  are the rightmost states in each row. Each additional $\hat B^\dagger$  operator moves us one step to the left (there are no more $\hat B^\dagger$ operators to apply after $m=-n+1$). All energies in a given row are degenerate. The dashed line shows where $E=0$.}
\label{fig: 2d_energy_schematic}
\end{figure}

At this stage, we show the spectra along with the operator representation of the radial components of the eigenvectors. All the states in a horizontal row (fixed $n$, $-n+1\le m\le n-1$) are degenerate with energy $E_{n-1}$. The column with $m=0$ has the largest number of eigenstates. As $|m|$ increases, the number of total eigenstates decreases by 1 for each step moving further away from $m=0$ (to the right or to the left). %We also indicate the operator form of the eigenvectors (operators acting on the $m=n-1$ state for the $n$th row).

Our next goal is to find the wavefunctions in position space and we now assume $m\ge 0$ again. As before, we see that the string of ${\hat B}^\dagger$ operators acting on the state $|n,n-1\rangle$ will create a polynomial of degree $n-m-1$ in $1/\hat\rho$. If we factor out  the term $1/\hat\rho^{n-m-1}$, then the polynomial is an order $n-m-1$ polynomial in $\hat\rho$. Just as we saw happen in three dimensions, it turns out to be a Laguerre polynomial for the two-dimensional case too. 

To see this, we start with a definition of the coefficients of the $n$th polynomial (for $m\ge 0$, without loss of generality):
\begin{eqnarray}
    &~&{\hat B}^\dagger_m\cdots {\hat B}_{n-2}^\dagger|n,n-1\rangle=\left (\frac{2i\hbar}{\sqrt{2\mu}\left ( n-\tfrac{1}{2}\right )a_0}\right )^{n-m-1}\\
    &~&~~\times\left (\frac{\left (n-\frac{1}{2}\right )a_0}{2\hat\rho}\right )^{n-m-1}\sum_{j=0}^{n-m-1}b_j^{(m)}\left (\frac{2\hat\rho}{\left (n-\frac{1}{2}\right )a_0}\right )^j|n,n-1\rangle.\nonumber
    \label{eq: 2d_polynomial_def}
\end{eqnarray}
Again, we suppress the index $n$ in the labeling of the $b$ coefficients for notational simplicity. We consider the base case first, corresponding to $m=n-2$. First, we need to determine a modification of the subsidiary condition in Eq.~(\ref{eq: 2d_subsidiary}) to determine what happens when the radial momentum acts on $|n,n-1\rangle$. It is
\begin{equation}
{\hat p}_\rho |n,n-1\rangle=i\hbar\left (
\frac{1}{\left (n-\frac{1}{2}\right )a_0}-\frac{n-\frac{1}{2}}{\hat\rho}\right )|n,n-1\rangle.
\label{eq: 2d_subsidiary2}
\end{equation}
Next, we have that the base case is
\begin{eqnarray}
&~&{\hat B}_{n-2}^\dagger|n,n-1\rangle\nonumber\\
&~&~~~=\frac{i\hbar}{\sqrt{2\mu}}\left ( \frac{2n-2}{\left (n-\frac{1}{2}\right )\left (n-\frac{3}{2}\right )a_0}-\frac{2n-2}{\hat\rho}\right )|n,n-1\rangle\nonumber\\
&~&~~~=\left (\frac{2i\hbar}{\sqrt{2\mu}\left (n-\frac{1}{2}\right )a_0}\right )\left (\frac{\left (n-\frac{1}{2}\right )a_0}{2\hat\rho}\right )\nonumber\\
&~&~~~~~~~~~\times \left ( -(2n-2)+\frac{2n-2}{2n-3}\frac{2\hat\rho}{\left (n-\frac{1}{2}\right )a_0}\right )|n,n-1\rangle,\nonumber\\
\label{eq: 2d_base_case}
\end{eqnarray}
so that $b_0^{(n-2)}=-(2n-2)$ and $b_1^{(n-2)}=(2n-2)/(2n-3)$.
This means $b_1^{(n-2)}/b_0^{(n-2)}=-1/(2n-3)$, a fact we shall use momentarily.

After establishing the base case for the proof by induction, we next examine the $m$th polynomial, by splitting off the operator ${\hat B}_m^\dagger$ and having it act on the $(m+1)$st polynomial. This yields
\begin{eqnarray}
&~&\sum_{j=0}^{n-m-1}\left (-(j+2m+2)b_j^{(m+1)}+\frac{m+n}{2m+1}b_{j-1}^{(m+1)}\right )\nonumber\\
&~&\times\left (\frac{2\hat\rho}{(n-\tfrac{1}{2})a_0}\right )^j=\sum_{j=0}^{n-m-1}b_j^{(m)}\left (\frac{2\hat \rho}{(n-\tfrac{1}{2})a_0}\right )^j.
\end{eqnarray}
Then we determine the ratio of subsequent coefficients of the polynomials, just like before and find that
\begin{equation}
    \frac{b_{j+1}^{(m)}}{b_j^{(m)}}=\frac{\scriptstyle{(j+2m+3)}\left (\frac{b_{j+1}^{(m+1)}}{b_j^{(m+1)}}\right )\scriptstyle{-}\frac{m+n}{2m+1}}{\scriptstyle{(j+2m+2)-}\frac{m+n}{2m+1}\left (\frac{b_j^{(m+1)}}{b_{j-1}^{(m+1)}}\right )^{-1}}~~\text{for}~~0\le j\le n-m-2,
    \label{eq: 2d_ratio}
\end{equation}
where we use again that $b_{-1}^{(m+1)}=b_{n-m-1}^{(m+1)}=0$.

The induction hypothesis is that
\begin{equation}
    \frac{b_{j+1}^{(m)}}{b_j^{(m)}}=\frac{j-n+m+1}{(j+1)(j+2m+1)}.
    \label{eq: 2d_induction_hypothesis}
\end{equation}
Recalling the base case, with $m=n-2$ and $j=0$, we have the ratio equals $-1/(2n-3)$, which is what we calculated above already for the base case. Substituting in the induction hypothesis for $m+1$ (evaluated for $j$ and for $j-1$) into Eq.~(\ref{eq: 2d_ratio}) yields
\begin{equation}
    \frac{b_{j+1}^{(m)}}{b_j^{(m)}}=\frac{\scriptstyle{(j-n+m+1)[(2m+1)(j-n+m+2)-(m+n)(j+1)]}}{\scriptstyle{(j+1)(j+2m+2)[(2m+1)(j-n+m+1)-(m+n)j]}}.
\end{equation}
Expanding the terms in the square brackets, factorizing them and simplifying, produces the results in Eq.~(\ref{eq: 2d_induction_hypothesis}), which completes the induction proof. Given the ratio of the consecutive coefficients in the Laguerre polynomial in Eq.~(\ref{eq: laguerre_ratio}), we see that the string of $B^\dagger$ operators acting on $|n,n-1\rangle$ yields the Laguerre polynomial (with $\alpha=2m$ and $m=n-m-1$) multiplied by a power of $\hat\rho$ multiplied by a constant, or
\begin{eqnarray}
    &~&{\hat B}^\dagger_m\cdots {\hat B}_{n-2}^\dagger|n,n-1\rangle=C'\left (\frac{2i\hbar}{\sqrt{2\mu}\left ( n-\tfrac{1}{2}\right )a_0}\right )^{n-m-1}\\
    &~&~~~\times\left (\frac{\left (n-\frac{1}{2}\right )a_0}{2\hat\rho}\right )^{n-m-1}L_{n-m-1}^{2m}\left (\frac{2\hat\rho}{\left ( n-\frac{1}{2}\right )a_0}\right )|n,n-1\rangle.\nonumber
\end{eqnarray}

Using the same approach as before, we compute the coefficient of the term that has no $\hat\rho$ dependence. Expressing the result in terms of the $b$ coefficients and equating it to the constant term found from using the subsidiary condition when the string of ${\hat B}^\dagger$ operators acts on $|n,n-1\rangle$. We have
\begin{eqnarray}
    &~&\left (\frac{2i\hbar}{\sqrt{2\mu}\left (n-\frac{1}{2}\right )a_0}\right )^{n-m-1}b_{n-m-1}^{(m)}\\
    &~&~~~=\left (\frac{i\hbar}{\sqrt{2\mu}}\right )^{n-m-1}\,\prod_{k=m}^{n-2}\left (\frac{1}{\left ( n-\frac{1}{2}\right )a_0}+\frac{1}{\left ( k+\frac{1}{2}\right ) a_0}\right ),\nonumber
\end{eqnarray}
which can be solved for $b_{n-m-1}^{(m)}$ with
\begin{equation}
    b_{n-m-1}^{(m)}=\frac{(2n-2)!(2m-1)!!}{(n+m-1)!(2n-3)!!}.
\end{equation}
We also can calculate this coefficient from the Laguerre polynomial form. We have
\begin{equation}
    b_{n-m-1}^{(m)}=C'a_{m-n-1}^{(2m,n-m-1)}=\frac{(-1)^{n-m-1}}{(n-m-1)!}.
\end{equation}
Hence, we have
\begin{equation}
    C'=(-1)^{n-m-1}\frac{(n-m-1)!(2n-2)!(2m-1)!!}{(n+m-1)!(2n-3)!!}
\end{equation}
and 
\begin{eqnarray}
|n,m\rangle&=&(-i)^{n-m-1}\sqrt{\frac{(n-m-1)!(2n-2)!}{(n+m-1)!}}\\
&\times&\left (\frac{\left (n-\frac{1}{2}\right )a_0}{2\hat\rho}\right )^{n-m-1}L_{n-m-1}^{2m}\left (\frac{2\hat\rho}{\left (n-\frac{1}{2}\right )a_0}\right )|n,n-1\rangle.\nonumber
\end{eqnarray}
As before, we drop the irrelevant phase factor for the remainder of the paper.

Now we are ready to compute the wavefunctions. We need to determine the overlap of $\langle \phi|\otimes\langle \rho|$ with $|n,m\rangle\otimes|m\rangle$. Since $\langle\phi|m\rangle=\exp(im\phi)/\sqrt{2\pi}$, and $\hat\rho|\rho\rangle=\rho|\rho\rangle$, we find that
\begin{eqnarray}
    \psi_{nm}(\rho)&=&\sqrt{\frac{(n-m-1)!(2n-2)!}{(n+m-1)!}}\langle\rho|n,n-1\rangle\\
    &\times&\left (\frac{\left (n-\frac{1}{2}\right )a_0}{2\rho}\right )^{n-m-1}L_{n-m-1}^{2m}\left (\frac{2\rho}{\left (n-\frac{1}{2}\right )a_0}\right ).\nonumber
\end{eqnarray}
This leaves us with the task of computing the wavefunction for maximal $m$ with a given $n$, that is with $m=n-1$. 

Here is where we need to use the translation operator in polar coordinates. We derived the translation operator when acting on the $|0_{\vec{r}}\rangle$ state. Hence, we have that
\begin{equation}
    \phi_n(\rho)=\langle\rho|n,n-1\rangle=\langle 0_\rho|\exp\left [ \frac{i\rho}{\hbar}\left ({\hat p}_\rho+\frac{i\hbar}{2\hat\rho}\right)\right ]|n,n-1\rangle.
\end{equation}
As before, the best way to evaluate this expression is to find the eigenvector of $\left ({\hat p}_\rho+\tfrac{i\hbar}{2\hat\rho}\right )$. Using Eq.~(\ref{eq: 2d_subsidiary2}), we see that
\begin{equation}
    {\hat p}_\rho \frac{1}{\hat\rho^{n-1}}|n,n-1\rangle=i\hbar\left (\frac{1}{\left (n-\frac{1}{2}\right )a_0}-\frac{1}{2\hat\rho}\right )\frac{1}{\hat\rho^{n-1}}|n,n-1\rangle,
\end{equation}
which leads to the eigenvalue/eigenvector relationship
\begin{equation}
    \left ({\hat p}_\rho+\frac{i\hbar}{2\hat\rho}\right )\frac{1}{\hat\rho^{n-1}}|n,n-1\rangle=\frac{i\hbar}{\left (n-\frac{1}{2}\right )a_0}\frac{1}{\hat\rho^{n-1}}|n,n-1\rangle.
\end{equation}
Hence, we have that
\begin{equation}
    \phi_n(\rho)=\langle 0_\rho|\frac{1}{\hat\rho^{n-1}}|n,n-1\rangle\rho^{n-1}\exp\left [-\frac{\rho}{\left (n-\frac{1}{2}\right )a_0}\right ].
\end{equation}
Normalizing gives us
\begin{equation}
    \langle 0_\rho|\frac{1}{\hat\rho^{n-1}}|n,n-1\rangle=\left (\frac{2}{\left (n-\frac{1}{2}\right )a_0}\right )^n\frac{1}{\sqrt{(2n-1)!}}.
\end{equation}
Summarizing, we find the normalized wavefunction is
\begin{eqnarray}
&~&\psi_{nm}(\rho,\phi)=\left (\frac{2}{\left (n-\frac{1}{2}\right )a_0}\right )^{|m|+1}\sqrt{\frac{(n-|m|-1)!}{(2n-1)(n+|m|-1)!}}\nonumber\\
&\times&\rho^{|m|}\exp\left [-\frac{\rho}{\left (n-\frac{1}{2}\right )a_0}\right ]L_{n-|m|-1}^{2|m|}\left (\frac{2\rho}{\left (n-\frac{1}{2}\right )a_0}\right )\frac{e^{im\phi}}{\sqrt{2\pi}}.\nonumber\\
\end{eqnarray}
This completes the calculation of the wavefunctions of the two-dimensional Coulomb problem.

\section{Conclusions}

In this paper, we have shown how one can calculate wavefunctions in a representation-independent manner.
The strategy starts from the origin in position space, translates to the position $\vec r$ and then computes the inner product with the energy eigenstate. Doing this for some central-potential problems, requires us to convert the translation operator from Cartesian coordinates to spherical or polar coordinates because a Taylor series expansion for the wavefunctions of some problems does not exist in terms of the Cartesian coordinates (this is true, for example, in the Coulomb problem). We illustrated how to convert from Cartesian coordinates to spherical (or polar) coordinates and then illustrated how the operator expression simplifies when it acts on the position-space eigenvector at the origin. Finally, we showed how such a procedure works by solving for the wavefunctions of the Coulomb problem in three and two dimensions. To do this requires us to use Schr\"odinger's factorization method, but we do so at the operator level only and do not convert the operators into the position-space representation. We want to emphasize that performing calculations in a representation-independent fashion illustrates that essentially all quantum properties are derived from the existence of  a few eigenvectors (such as the position-space eigenvector at the origin and the ground-state eigenvector of the Coulomb problems) and the canonical commutation relation $[{\hat r}_\alpha,{\hat p}_\beta]=i\hbar\delta_{\alpha\beta}$. No other assumptions are needed. This methodology is quite general, but it is simpler for position-space wavefunctions than momentum-space wavefunctions. This is because the generic factorization method has raising and lowering operators that are linear in momentum operators, but often are nonlinear in the position operators. Hence, it is easier to determine how the momentum operator acts on an energy eigenstate than it is to determine how a position operator acts. Nevertheless, this approach can be used in both position and momentum space and explicitly shows that these two wavefunctions are constructed in similar ways. The approach to determine them from ``differential equations'' is often quite different, as the general formulation in momentum space is an integral equation rather than a differential equation, which is what is used in position space.

\acknowledgments

This work was supported by the National Science Foundation under grant number PHY-1915130. In addition, JKF was supported by the McDevitt bequest at Georgetown University. We also acknowledge useful discussions with Xinliang Lyu and Christina Daniel. Data sharing policy: Data sharing not applicable---no new data generated.

\appendix*

\section{Representation-independent calculation of commutators}

In the early years of quantum mechanics, Wolfgang Pauli~\cite{pauli} and Paul Dirac~\cite{dirac} showed how to use the canonical commutation relation to compute commutators of functions of $\hat r_\alpha$ and $\hat p_\alpha$. We use this same method to compute the commutators needed in this work. To start, we compute the commutator of momentum with $\hat r^2={\hat r}_x^2+{\hat r}_y^2+{\hat r}_z^2$. Using the Leibnitz rule for the commutator of a product, we immediately find
\begin{equation}
    [{\hat p}_\alpha,\hat r^2]= {\hat r}_\alpha[{\hat p}_\alpha,{\hat r}_\alpha]+[{\hat p}_\alpha,{\hat r}_\alpha]{\hat r}_\alpha=-2i\hbar {\hat r}_\alpha;
    \label{eq: rsquared}
\end{equation}
note that we never use the Einstein summation convention in this paper.
Similarly, defining ${\hat r}=\sqrt{{\hat r}_x^2+{\hat r}_y^2+{\hat r}_z^2}$, the Leibnitz rule shows that
\begin{equation}
    [{\hat p}_\alpha,\hat r^2]=\hat r[{\hat p}_\alpha,\hat r]+[{\hat p}_\alpha,\hat r]\hat r=-2i\hbar {\hat r}_\alpha,
    \label{eq: rsquared2}
\end{equation}
after using the result in Eq.~(\ref{eq: rsquared}).
The commutator $[{\hat p}_\alpha,{\hat r}]$ commutes with $\hat r$ as described by B\"ohm~\cite{bohm}, which follows from the triple commutator with  $\hat r^2$
\begin{equation}
    \big [\hat r^2,[{\hat p}_\alpha,\hat r]\big ]=\hat r\big [\hat r,[{\hat p}_\alpha,\hat r]\big ]+\big [\hat r,[{\hat p}_\alpha,\hat r]\big ]\hat r
    \label{eq: triple_commutator}
\end{equation}
and moving the $\hat r$ operator into the second element of the outermost commutator to multiply the term $[{\hat p}_\alpha,\hat r]$ for both nested commutators (which is valid because $[\hat r,\hat r]=0$) yields
\begin{equation}
     \big [\hat r,\hat r[{\hat p}_\alpha,\hat r]+[{\hat p}_\alpha,\hat r]\hat r\big ]=[\hat r,[{\hat p}_\alpha,\hat r^2]]=[\hat r,(-2i\hbar {\hat r}_\alpha)]=0
     \label{eq: nested_commutator}
\end{equation}
after substituting Eq.~(\ref{eq: rsquared}) into the innermost commutator. Because the square root of an operator is uniquely defined to have the same eigenvectors as the original operator (but all eigenvalues are equal to the positive square roots of the eigenvalues of the original operators), if an operator $\hat A$ commutes with another operator $\hat B$, then the square root of $\hat A$, also commutes with $\hat B$. Hence, from Eq.~(\ref{eq: nested_commutator}), we have $\big [\hat r,[{\hat p}_\alpha,\hat r]\big ]=0$. Combining this with Eq.~(\ref{eq: rsquared2}) gives us
\begin{equation}
    [\hat p_\alpha,\hat r]=-i\hbar\frac{{\hat r}_\alpha}{\hat r}.
    \label{eq: momentum_r_comm}
\end{equation}
One uses the Leibnitz rule again to compute the commutator with $1/\hat r$ via
\begin{equation}
    0=[{\hat p}_\alpha,1]=\left [{\hat p}_\alpha,\frac{\hat r}{\hat r}\right ]=-i\hbar\frac{{\hat r}_\alpha}{\hat r^2}+\hat r\left [{\hat p}_\alpha,\frac{1}{\hat r}\right ],
    \label{eq: comm_trick}
\end{equation}
which can be re-arranged to give
\begin{equation}
    \left [{\hat p}_\alpha,\frac{1}{\hat r}\right ]=i\hbar\frac{{\hat r}_\alpha}{\hat r^3}.
    \label{eq: momentum_1/r_comm}
\end{equation}

Similarly, using $\hat\rho=\sqrt{{\hat r}_x^2+{\hat r}_y^2}$, we find that $[{\hat p}_\alpha,\hat\rho]=-i\hbar{\hat r}_\alpha/\hat \rho$ for $\alpha=x$ or $y$ and it vanishes for $\alpha=z$. We also find that $[{\hat p}_\alpha,1/\hat\rho]=i\hbar{\hat  r}_\alpha/{\hat\rho}^3$ with the same conditions on $\alpha$.

Similarly, we define $\cos\hat\theta={\hat r}_z/\hat r$, $\sin\hat\theta=\hat\rho/\hat r$, $\cos\hat\phi={\hat r}_x/\hat\rho$, and $\sin\hat\phi={\hat r}_y/\hat\rho$.
Then using the Leibnitz rule, we find that 
\begin{equation}
    [{\hat p_x},\cos\hat\theta]=i\hbar\frac{{\hat r}_x{\hat r}_z}{\hat r^3},~~[{\hat p_y},\cos\hat\theta]=i\hbar\frac{{\hat r}_y{\hat r}_z}{\hat r^3},~~[{\hat p_z},\cos\hat\theta]=-i\hbar\frac{{\hat \rho}^2}{\hat r^3},
    \label{eq: momentum_costheta_comm}
\end{equation}
\begin{equation}
    [{\hat p_x},\sin\hat\theta]=-i\hbar\frac{{\hat r}_x{\hat r}_z^2}{\hat\rho\hat r^3},~~[{\hat p_y},\sin\hat\theta]=-i\hbar\frac{{\hat r}_y{\hat r}_z^2}{\hat\rho\hat r^3},~~[{\hat p_z},\sin\hat\theta]=i\hbar\frac{{\hat \rho}{\hat r}_z}{\hat r^3},
    \label{eq: momentum_sintheta_comm}
\end{equation}
\begin{equation}
    [{\hat p_x},\cos\hat\phi]=-i\hbar\frac{{\hat r}_y^2}{\hat \rho^3},~~[{\hat p_y},\cos\hat\phi]=i\hbar\frac{{\hat r}_x{\hat r}_y}{\hat \rho^3},~~[{\hat p_z},\cos\hat\phi]=0,
    \label{eq: momentum_cosphi_comm}
\end{equation}
and
\begin{equation}
    [{\hat p_x},\sin\hat\phi]=i\hbar\frac{{\hat r}_x{\hat r}_y}{\hat \rho^3},~~[{\hat p_y},\sin\hat\phi]=-i\hbar\frac{{\hat r}_x^2}{\hat \rho^3},~~[{\hat p_z},\sin\hat\phi]=0.
    \label{eq: momentum_sinphi_comm}
\end{equation}
Of course these ``trigonometric function'' operators commute with themselves, and with $\hat \rho$ and $\hat r$, because all position operators commute with each other. Note that the angle operators are well defined {\it only} when they are arguments of the trigonometric functions. We never work with angle operators by themselves.

As described in the main text, we use the unit-vector operators (${\hat{\vec e}}_r=\sin\hat\theta\cos\hat\phi {\vec e}_x+\sin\hat\theta\sin\hat\phi{\vec e}_y+\cos\hat\theta{\vec e}_z$, ${\hat{\vec e}}_\theta=\cos\hat\theta \cos\hat\phi\vec{e}_x+\cos\hat\theta\sin\hat\phi\vec{e}_y-\sin\hat\theta {\vec e}_z$ and ${\hat{\vec e}}_\phi=-\sin\hat\phi {\vec e}_x+\cos\hat\phi{\vec e}_y$) to define the components of momentum in spherical coordinates; in spherical coordinates, the unit vectors must be operators, while in Cartesian space, they are not. Using the symmetric combination of the dot product of momentum with these unit vectors (given schematically as $\tfrac{1}{2}{\hat{\vec e}}\cdot{\hat{\vec p}}+\tfrac{1}{2}{\hat{\vec p}}\cdot{\hat{\vec e}}$) and after computing the ``quantum corrections'' (proportional to $i\hbar$), we find that
\begin{equation}
    {\hat p}_r=\sin\hat\theta\cos\hat\phi{\hat p}_x+\sin\hat\theta\sin\hat\phi{\hat p}_x+\cos\hat\theta{\hat p}_z-\frac{i\hbar}{\hat r},
    \label{eq: radial_mom}
\end{equation}
\begin{equation}
    {\hat p}_\theta=\cos\hat\theta\cos\hat\phi{\hat p}_x+\cos\hat\theta\sin\hat\phi{\hat p}_y-\sin\hat\theta{\hat p}_z-i\hbar\frac{\text{cot}\hat\theta}{2\hat r},
    \label{eq: theta_mom}
\end{equation}
and
\begin{equation}
    {\hat p}_\phi=-\sin\hat\phi{\hat p}_x+\cos\hat\phi{\hat p}_y.
    \label{eq: phi_mom}
\end{equation}
Note that ${\hat p}_\phi$ does not have a quantum correction due to reordering. This is because it is equal to ${\hat L}_z/ \hat\rho$.

Now that we have the components of momentum along the different spherical coordinate directions, we can use the commutation relations of the Cartesian components of momentum with the radial and angular operators to find that
\begin{equation}
    [{\hat p}_r,\hat r]=-i\hbar,~~[{\hat p}_r,\hat\rho]=-i\hbar\sin\hat\theta
    \label{eq: radial_momentum_r_comm}
\end{equation}
and the commutator of ${\hat p}_r$ with any trigonometric function of  angles $\hat\theta$ and $\hat\phi$ is zero. We also have that ${\hat p}_\theta$ commutes with functions of $\hat r$ and $\hat\phi$. But, we have
\begin{equation}
    [{\hat p}_\theta,\cos\hat\theta]=i\hbar\frac{\sin\hat\theta}{\hat r}~~\text{and}~~[{\hat p}_\theta,\sin\hat\theta]=-i\hbar\frac{\cos\hat\theta}{\hat r}.
    \label{eq: theta_momentum_trig_comm}
\end{equation}
In addition, we have
\begin{equation}
    [{\hat p}_\theta,\hat\rho]=-i\hbar\cos\hat\theta.
    \label{eq: theta_mom_rho_comm}
\end{equation}
Finally, for ${\hat p}_\phi$, we find it commutes with trigonometric functions of $\hat\theta$ and arbitrary functions of $\hat r$ and $\hat\rho$. We also find that
\begin{equation}
    [{\hat p}_\phi,\cos\hat\phi]=i\hbar\frac{\sin\hat\phi}{\hat\rho}~~\text{and}~~[{\hat p}_\phi,\sin\hat\phi]=-i\hbar\frac{\cos\hat\phi}{\hat\rho}.
    \label{eq: phi_momentum_trig_com}
\end{equation}

Surprisingly, because we have projected the momentum onto components of unit-vector operators, which are not the canonical momenta, we need to compute their commutators, which do not vanish, in general. We find that
\begin{equation}
    [{\hat p}_r,{\hat p}_\theta]=\frac{i\hbar}{\hat r}{\hat p}_\theta,
    \label{eq: radial_momentum_theta_momenum_comm}
\end{equation}
\begin{equation}
    [{\hat p}_r,{\hat p}_\phi]=\frac{i\hbar}{\hat r}{\hat p}_\phi,
    \label{eq: radial_momentum_phi_momenum_comm}
\end{equation}
and
\begin{equation}
   [{\hat p}_\theta,{\hat p}_\phi]= i\hbar\frac{\text{cot}{\hat\theta}}{\hat r}{\hat p}_\phi.
   \label{eq: theta_momentum_phi_momenum_comm}
\end{equation}

In two dimensions, there are only a few changes. We define ${\hat{\vec e}}_\rho=({\hat r}_x{\vec e}_x+{\hat r}_y{\vec e}_y)/\hat\rho$ and ${\hat {\vec e}}_\phi=(-{\hat r}_y{\vec e}_x+{\hat r}_x{\vec e}_y)/\hat\rho$, the latter vector being the same as in three dimensions, when defined in terms of Cartesian operators. using the trigonometric operators, these become 
\begin{equation}
    {\hat p}_\rho=\cos\hat\phi {\hat p}_x+\sin\hat\phi{\hat p}_y- \frac{i\hbar}{2\hat \rho}
    \label{eq: radial_momentum_2d}
\end{equation}
    and 
\begin{equation} 
    {\hat p}_\phi=-\sin\hat\phi{\hat p}_x+\cos\hat\phi{\hat p}_y
    \label{eq: phi_momentum_2d}
\end{equation}
after evaluating the quantum correction for the radial momentum. The commutators are similar and we report them here. First the radial momentum, which satisfies
\begin{equation}
    [{\hat p}_\rho,\hat\rho]=-i\hbar,
    \label{eq: radial_momentum_rho_com_2d}
\end{equation}
while ${\hat p}_\rho$ commutes with $\cos\hat\phi$ and $\sin\hat\phi$. Next the $\phi$-component of the momentum
\begin{equation}
    [{\hat p}_\phi, \cos\hat\phi]=i\hbar\frac{\sin\hat\phi}{\hat\rho}~~\text{and}~~[{\hat p}_\phi,\sin\hat\phi]=-i\hbar\frac{\cos\hat\phi}{\hat\rho}.
      \label{eq: phi_momentum_trig_com_2d}
\end{equation}
We also have $[{\hat p}_{\phi},\hat\rho]=0$. Finally, we compute the commutation relation between the components of momentum. This yields
\begin{equation}
    [{\hat p}_\rho,{\hat p}_\phi]=\frac{i\hbar}{\hat\rho}{\hat p}_\phi.
      \label{eq: radial_momentum_phi_momentum_comm_2d}
\end{equation}

The last set of identities we derive in this appendix is the conversion of the kinetic-energy operator into its radial and angular components. Beginning in three dimensions, we find that
\begin{eqnarray}
    &~&{\hat p}_r^2=\frac{1}{\hat r}({\hat r}_x{\hat p}_x+{\hat r}_y{\hat p}_y+{\hat r}_z{\hat p}_z-i\hbar)\frac{1}{\hat r}({\hat r}_x{\hat p}_x+{\hat r}_y{\hat p}_y+{\hat r}_z{\hat p}_z-i\hbar)\nonumber\\
    &~&~~~\frac{1}{\hat r^2}[{\hat r}_x^2{\hat p}_x^2+{\hat r}_y^2{\hat p}_y^2+{\hat r}_z^2{\hat p}_z^2\nonumber\\
   &~&~~~~~~~+2{\hat r}_x{\hat r}_y{\hat p}_x{\hat p}_y+
   2{\hat r}_y{\hat r}_z{\hat p}_y{\hat p}_z+
   2{\hat r}_z{\hat r}_x{\hat p}_x{\hat p}_x\nonumber\\
   &~&~~~~~~~
   -2i\hbar({\hat r}_x{\hat p}_x+{\hat r}_y{\hat p}_y+{\hat r}_z{\hat p}_z)]
   \label{eq: radial_momentum_squared}
\end{eqnarray}
Next, using the standard definition of orbital angular momentum $L_\alpha=\sum_{\beta\gamma}\epsilon_{\alpha\beta\gamma}{\hat r}_\beta{\hat p}_\gamma$, we find that
\begin{eqnarray}
    &~&\frac{1}{{\hat r}^2}{\hat{\vec L}}\cdot{\hat{\vec L}}=\frac{1}{\hat r^2}[({\hat r}_x{\hat p}_y-{\hat r}_y{\hat p}_x)({\hat r}_x{\hat p}_y-{\hat r}_y{\hat p}_x)\nonumber\\
   &~ &~~~+({\hat r}_y{\hat p}_z-{\hat r}_z{\hat p}_y)({\hat r}_y{\hat p}_z-{\hat r}_z{\hat p}_y)+({\hat r}_z{\hat p}_x-{\hat r}_x{\hat p}_z)({\hat r}_z{\hat p}_x-{\hat r}_x{\hat p}_z)]\nonumber\\
   &~&~~~=\frac{1}{\hat r^2}[({\hat r}_y^2+{\hat r}_z^2){\hat p}_x^2+({\hat r}_x^2+{\hat r}_z^2){\hat p}_y^2+({\hat r}_x^2+{\hat r}_y^2){\hat p}_z^2\nonumber\\
   &~&~~~~~~~-2{\hat r}_x{\hat r}_y{\hat p}_x{\hat p}_y-
   2{\hat r}_y{\hat r}_z{\hat p}_y{\hat p}_z-
   2{\hat r}_z{\hat r}_x{\hat p}_x{\hat p}_x\nonumber\\
   &~&~~~~~~~
   +2i\hbar({\hat r}_x{\hat p}_x+{\hat r}_y{\hat p}_y+{\hat r}_z{\hat p}_z)].
   \label{eq: angular_momentum_squared}
\end{eqnarray}
Adding them together yields
\begin{equation}
    \frac{1}{2\mu}({\hat p}_x^2+{\hat p}_y^2+{\hat p}_z^2)=\frac{1}{2\mu}\left ({\hat p}_r^2+\frac{{\hat{\vec L}}\cdot{\hat{\vec L}}}{\hat r^2}\right ).
    \label{eq: ke_3d}
\end{equation}
Note that there are no quantum corrections in this case.

In two dimensions, we find that
\begin{eqnarray}
{\hat p}_\rho^2&=&\frac{1}{\hat \rho}({\hat r}_x{\hat p}_x+{\hat r}_y{\hat p}_y-\tfrac{i}{2}\hbar)\frac{1}{\hat\rho}({\hat r}_x{\hat p}_x+{\hat r}_y{\hat p}_y-\tfrac{i}{2}\hbar)\\
&=&\frac{1}{\hat\rho^2}\left [{\hat r}_x^2{\hat p}_x^2+{\hat r}_y^2{\hat p}_y^2+2{\hat r}_x{\hat r}_y{\hat p}_x{\hat p}_y-i\hbar({\hat r}_x{\hat p}_x+{\hat r}_y{\hat p}_y)+\frac{\hbar^2}{4}\right ]\nonumber
\label{eq: radial_momentum_squared_2d}
\end{eqnarray}
and
\begin{equation}
    \frac{1}{\hat\rho^2}{\hat L}_z^2=\frac{1}{\hat \rho^2}[{\hat r}_y^2{\hat p}_x^2+{\hat r}_x^2{\hat p}_y^2-2{\hat r}_x{\hat r}_y{\hat p}_x{\hat p}_y+i\hbar({\hat r}_x{\hat p}_x+{\hat r}_y{\hat p}_y)].\\
    \label{eq: phi_momentum_squared_2d}
\end{equation}
Again, adding these two together yields
\begin{equation}
    \frac{1}{2\mu}({\hat p}_x^2+{\hat p}_y^2)=\frac{1}{2\mu}\left ({\hat p}_\rho^2+\frac{{\hat L}_z^2-\tfrac{1}{4}\hbar^2}{\hat\rho^2}\right ).
    \label{eq: ke_2d}
\end{equation}
In this case, there is a quantum correction (the term proportional to $\hbar^2$)!

\end{document}